\documentclass[useAMS,usenatbib]{mnras}
\usepackage{graphicx}
\usepackage{natbib}

\title[Sculpting the MW disc with Sgr and the LMC]{The Influence of Sagittarius and the Large Magellanic Cloud on the Milky Way Galaxy}

\author[Laporte et al.]{
\parbox[t]{\textwidth}{Chervin F. P. Laporte$^{1}$\thanks{$\!\!$Simons Fellow   e-mail:cfl2126@columbia.edu}, Kathryn V. Johnston$^{1}$, Facundo A. G\'omez$^{2,3,4}$, Nicolas Garavito-Camargo$^{5}$ \& Gurtina Besla$^{5}$}\\
$^{1}$ Department of Astronomy, Columbia University, 550 West 120th Street, New York, NY, 10027, U.S.A\\
$^{2}$ Instituto de Investigaci\'on Multidisciplinar en Ciencia y Tecnolog\'ia, Universidad de La Serena, Ra\'ul Bitr\'an 1305, La Serena, Chile\\ 
$^{3}$ Departamento de F\'isica y Astronom\'ia, Universidad de La Serena, Norte, Av. Juan Cisternas 1200, La Serena, Chile\\
$^{4}$ Max-Planck-Institut f{\"u}er Astrophysik, Karl-Schwarzschild-Str. 1, D-85748, Garching bei M{\"u}enchen, Germany\\
$^{5}$ Steward Observatory, University of Arizona, 933 North Cherry Avenue, Tucson, AZ, 8572, U.S.A.\\
}
\begin{document}
\date{}

\pagerange{\pageref{firstpage}--\pageref{lastpage}} \pubyear{2011}
\maketitle
\label{firstpage}
\begin{abstract}
%following the interaction all the way from the time when Sgr first crosses the MW's viral radius to the present day

We present N-body simulations of a Sagittarius-like dwarf spheroidal galaxy (Sgr) that follow its orbit about the Milky Way (MW) since its first crossing of the Galaxy's virial radius to the present day. As Sgr orbits around the MW, it excites vertical oscillations, corrugating and flaring the Galactic stellar disc. These responses can be understood by a two-phase picture in which the interaction is first dominated by torques from the wake excited by Sgr in the MW dark halo before transitioning to  tides from Sgr's direct impact on the disc at late times. We show for the first time that a massive Sgr model simultaneously reproduces the locations and motions of arc-like over densities, such as the Monoceros Ring and the Triangulum Andromeda stellar clouds, that have been observed at the extremities of the disc, while also satisfying the solar neighbourhood constraints on the vertical structure and streaming motions of the disc. In additional simulations, we include the Large Magellanic Cloud (LMC) self consistently with Sgr. The LMC introduces coupling through constructive and destructive interference, but no new corrugations. In our models, the excitation of the current structure of the outer disk can be traced to interactions as far back as 6-7 Gyr ago (corresponding to $z\leq1$). Given the apparently quiescent accretion history of the MW over this timescale, this places Sgr as the main culprit behind the vertical oscillations of the disc and the last major accretion event for the Galaxy with the capacity to modulate its chemodynamical structure.

\end{abstract}

\section{Introduction}

The existence and age of the thin stellar disc in our Galaxy implies that it has not suffered a major dynamical interaction for many Gigayears, and hence the Milky Way (MW) is often thought to be a good laboratory for studying the steady, secular evolution of galaxies in isolation \citep{freeman02}. However, there is increasing observational evidence that the Galaxy is still experiencing quite a tumultuous and exciting life. The presence of a smooth stellar halo \citep{deason11a} possibly indicates that the MW may have had a more brutal past, the signatures of which were dynamically mixed during episodes of violent relaxation \citep{Bullock2005}. However, a major part of the stellar halo is locked into debris from the Sagittarius (hereafter Sgr) dwarf spheroidal \citep{bell08}. Sgr is currently on its way to full disruption, with streams of debris clearly mapped out to large galactocentric radii over a large fraction of the sky  \citep{majewski03,belokurov06, koposov12, sesar17}. These streams wrap more than once entirely around our Galaxy, demonstrating that Sgr must have plunged a few times through the disc and may have impacted its structure both in the solar neighbourhood but also at large galactocentric radii. The importance of the influence of this dwarf companion on  the Galaxy is not yet clear since the progenitor mass of Sgr is still not well constrained. \cite{jiang00} were amongst the first theoretical studies addressing the range of possible infall masses for the Sgr dSph compatible with its current position and velocity vector. They went as far as to propose that the satellite might have been as massive as $10^{11}\,\rm{M_{\odot}}$, a possibility that has not been explored further when typically studying the stream of Sgr \citep{helmi01,helmi04,johnston05,penarrubia10, law10}, until just recently \citep{purcell11,gibbons16}. In addition to the Sgr dwarf, there is increasing evidence that the Magellanic Clouds are on their first infall onto the Milky Way \citep{besla07,kallivayalil13}. This opens the door for the clouds to be more massive than previously thought \citep{boylan-kolchin11, penarrubia16} which would have important repercussions on the structure of the disc \citep{weinberg06,laporte16} as well, particularly in exciting the warp traced by the neutral HI gas. Currently, no comprehensive study of the MW major accretion history events have ever been carried (i.e. Sgr and Magellanic Clouds).

A mass of $10^{11}\,\rm{M_{\odot}}$ would put Sgr in the range of minor merger events which are known to modulate the structure of Galactic discs in several ways. They can excite spiral arms and bar formation \citep{purcell11}, dynamically cold ring-like features \citep{younger08}, thicken pre-existing thin discs \citep{velazquez99, villalobos08, purcell10}, cause ringing in the $u-v$ and $E-Lz$ planes \citep{minchev09,gomez12} or produce vertical oscillations \citep{gomez13,widrow14,gomez16} which are coupled with motions in the radial direction \citep{d'onghia16}. 

Recent surveys have uncovered indications of vertical oscillations in and around our Galactic stellar disc. In the solar neighbourhood, \cite{widrow12} identified a north-south asymmetry in vertical velocity and also in space through residual number density estimates. Vertical streaming motions are also seen in the LAMOST survey \citep{carlin13} and RAVE data \citep{williams13}. Although, the existence of some of these kinematical asymmetries may also be tied to interactions with the bar and spiral arms \citep{monari15,monari16} or their coupling \citep{monari16b}, these do not exclude a minor merger driven origin. However, to our knowledge, the bar or spiral arms {\it in isolation cannot} give rise to asymmetries in physical space such as those observed by \cite{widrow12}.

Various photometric studies have also identified possible signatures of vertical oscillations well beyond the solar neighbourhood, apparent as low-latitutde overdensities in stars counts above and below the plane of the disc. A poster-child example of this is the Monoceros Ring also known as the Galactic Anticenter Stellar Structure (GASS) overdensity lying at heliocentric distances $5\leq d_{\odot}/{\rm{kpc}}\leq13$ towards the Anticenter  \citep{newberg03,ibata03,crane03}. This structure as recently been mapped by the $3\pi$ Pan-STARRS survey \citep{slater14, morganson16}. The Monoceros Ring spans a large area in the sky with Galactic longitudes $120^{\circ}\leq l\leq240^{\circ}$ and latitudes $-30^{\circ}\leq b \leq +40^{\circ}$. Signs of oscillations in the direction of Triangulum and Andromeda \citep{martin07} have also been mapped by \cite{xu15} as well as in A13 \citep{li17}. TriAnd lies at a Galactocentric distance of $R\sim25-30\,\rm{kpc}$ at $Z\sim -10\,\rm{kpc}$ below the mid-plane around $160\leq l \leq120^{\circ}$ \citep{sheffield14} and has stellar populations consistent with it being part of the disc \citep{price-whelan15}. A13 \citep{sharma10} is another association of M giants at an intermediate distance just between GASS and TriAnd but located in the Nothern Galactic Hemisphere in the area $125^{\circ}\leq l \leq210^{\circ}$, $+20^{\circ}\leq b \leq +40^{\circ}$ at a heliocentric distance of $d_{\odot}\sim 10-20\,\rm{kpc}$ also thought to be part of the disc \citep{li17}. Finally, in the direction of the Galactic Center ($l\sim0^{\circ}$), \cite{feast14} have identified Cepheids at $R\sim15-22 \,\rm{kpc}$ which lie above and below the midplane at absolute heights of $|Z|\sim2\,\rm{kpc}$ suggesting that they could be part of the flared disc, or possibly as we shall see, a result of vertical oscillations.

Different theoretical explanations for the formation of low latitude overdensities have been put forward. These have included the tidal interaction of multiple mergers \citep{kazantzidis08}, a single event encounter with a dwarf galaxy \citep{younger08}, a low-velocity fly-by \citep{gomez15b} or the tidal disruption of satellites in the plane of the disc \citep{penarrubia05,sheffield14}. An in-plane accretion origin seems difficult to reconcile with the fact that these structures exhibit disc-like motions \citep{deboer17}. Additionally, there is now clear mounting evidence that the stellar populations of these features are consistent with them originating from the disc (Sheffield et al. submitted, Bergemann et al. submitted). Furthermore, the multiple merger scenario although pertinent may be hard to reconcile given the relatively quiet accretion history of the Milky Way \citep{ruchti15} and the fact that Sgr dominates the stellar halo mass budget \citep{bell08,ostholt10}. Given that Sgr has undergone multiple passages through the disc, \cite{dansac11} have hypothesized that the Sgr dwarf spheroidal may be responsible for exciting the formation of the Monoceros Ring as its past orbit crosses at a similar radius to the location of the Monoceros Ring. This idea was tested in \cite{purcell11} through live N-body simulations where they were able to show qualitative matching behaviour. \cite{gomez13} used those simulations to show that Sgr can excite oscillatory features in velocity and physical space that are qualitatively similar to those observed in \cite{widrow12} but with different amplitudes and wavelengths. Using the same initial conditions from \cite{purcell11} but extending the disc out to $R\sim40 \,\rm{kpc}$, \cite{price-whelan15} used the simulations of \cite{laporte16} to show that the vertical waves identified in \cite{gomez13} can excite disc material below the plane of the disc around the galactocentric radii and phases were the TriAnd clouds are observed. 

However, none of these models were in strict quantitative agreement with the observations in terms of the vertical heights reached by the kicked-up stars. We suspect that part of the problem in these earlier N-body experiments following the response of the MW disc to Sgr is that they all assumed initially stable discs and artificially truncated satellites set on orbits well within the virial radius of the MW \citep{purcell11,gomez13,laporte16}, thus neglecting multiple earlier passages. It is well known from linear perturbation theory that as satellites sink within their host via dynamical friction, these excite wakes in the host dark matter halo of a galaxy which can warp discs \citep{weinberg89}. This mechanism, has been shown to be particularly relevant to the Milky Way as it can explain the shape of the HI gas warp \citep{weinberg06, laporte16} which forms as the LMC excites a wake in the MW dark matter halo. We suspect that this mechanism could equally be important for seeding earlier perturbations in the stellar disc mediated by Sgr's multiple pericentric passages. In fact, a satellite may not need to be very massive or at close pericentric passage in order to produce wakes inside its host halo \citep{vesperini00}. Indeed, \cite{gomez15b} used a cosmological hydrodynamical simulation of a MW-type galaxy to show the importance of low-velocity fly-bys on the vertical structure of Galactic stellar discs and how satellites with large pericentric radii can cause a dark matter halo wake inside a galaxy to give rise to Monceros-like features. Moreover, in \cite{laporte16} we hypothesised that the coupling between the Sgr dSph and a massive LMC could potentially affect the vertical structure of the disc in a non-negligible way as we showed that each galaxy on its own is capable of inflicting disturbances of the same order of magnitude\footnote{The nature of the perturbations from each satellite are distinct. Because the LMC is just past first pericenter, it only produces a warp whereas Sgr due to its multiple passages through the disc produces corrugations in the stellar disc.}. We suggested this could possibly explaining some of the discrepancies between the shape of the observed HI warp of the Milky Way and those of our earlier models \footnote{It should be noted that the inferences from the shape of the HI warp are based on kinematic distance estimates which are known to be severely biased by up to a factor of two \citep{reid14}.}. In this work, our simulations are collisionless so our attention will be primarily focused on the structure of the stellar disc - any allusion to the cold gas HI warp should be interpreted in a strict loose sense and with caution as collisional gas dynamics is fundamentally different to that of stars.
%can explain the existence of the MW's HI warp \citep{weinberg06, laporte16} as the MW halo dark halo responds to the LMC's passage within it. 

In this paper we explore whether Sgr alone can plausibly be responsible for exciting the vertical oscillations of the stellar disc apparent in and around our Galaxy, extending prior work to include the entirety of the dynamical interaction starting out to the virial radius. The challenge is whether such a model for Sgr may exist such that it is capable to reproduce low-latitude features towards the Anticenter from the Monoceros Ring out to TriAnd. This is illustrated in Figure 1, where we present a stellar density map of main sequence turn-off stars taken from \cite{slater14} overplotted with stars associated with GASS \citep{crane03}, A13 \citep{li17} and TriAnd \citep{sheffield14} which are situated behind Monoceros and extend out to extreme Galactocentric distances $R\sim25-30 \,\rm{kpc}$ and heights $Z\sim-10\,\rm{kpc}$.

\begin{figure*}
\includegraphics[width=1.0\textwidth,trim=50mm 50mm 40mm 50mm, clip]{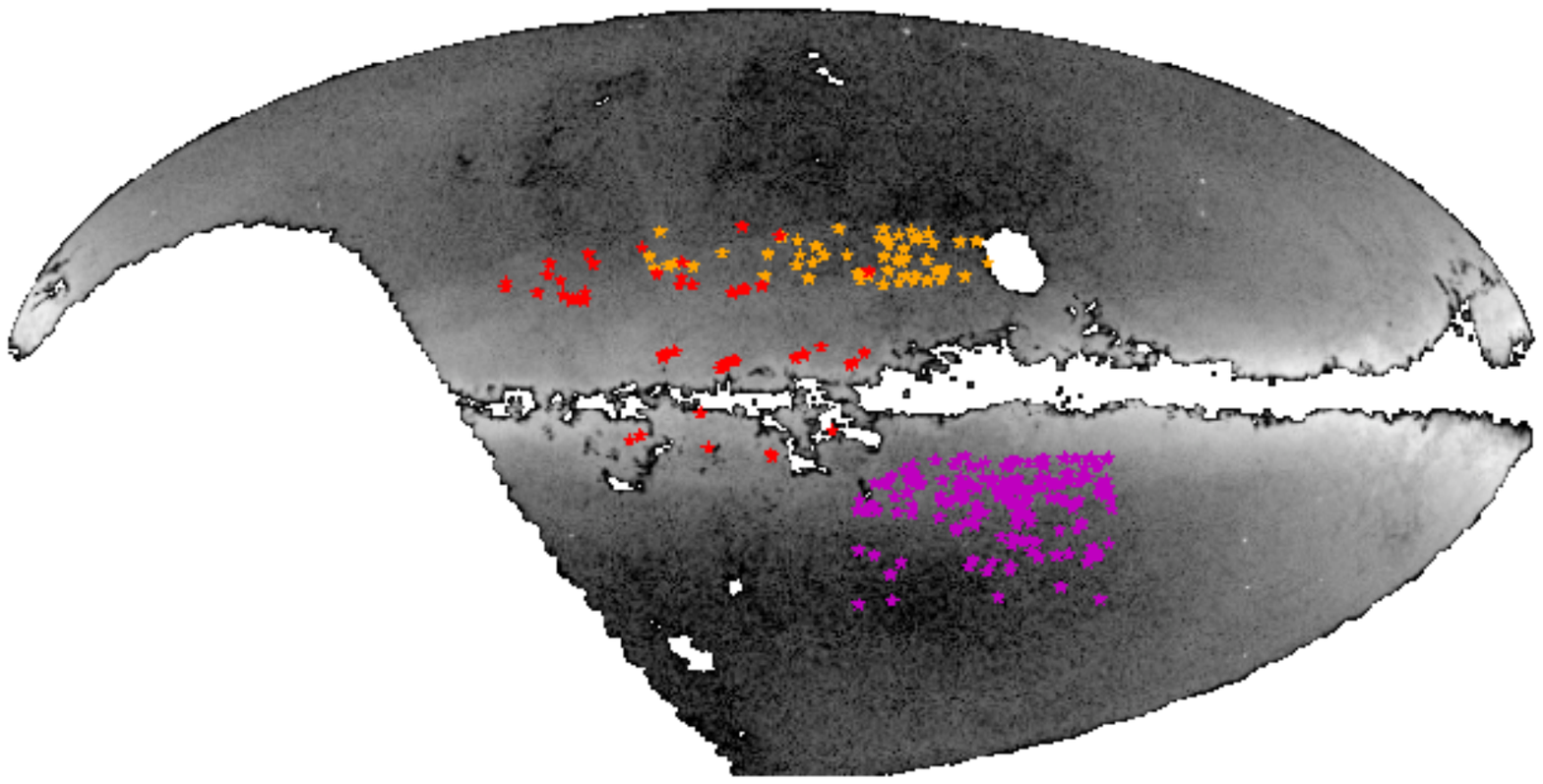}
\caption[]{Stellar density map of main sequence turn-off stars from the Pan-STARRS catalog of \cite{slater14} at a heliocentric distance of $\sim10\,\rm{kpc}$. Monoceros is clearly visible at those distances. Overplotted are GASS stars from \cite{crane03}, the A13 stars from \cite{li17} and TriAnd stars from \cite{sheffield14} (red, orange, magenta filled stars respectively). The vertical stream perpendicular to the midplane in the Anticenter is belongs to Sgr. This figure is meant to give a summary illustration of the low-latitude features observed in the vicinity of this disc. We see that the A13 stars which are further away from Monoceros, delineate quite well the stellar density map from Pan-STARRS as noted earlier by \cite{li17}. The TriAnd stars are much further away in the disc around galactocentric distances close to $R=25-30\,\rm{kpc}$ and heights corresponding to $Z=-10 \,\rm{kpc}$. One of the aims of this paper is to test whether all these features can all be reconciled under a single accretion event, specifically Sgr. A similar figure to this was already presented in \cite{li17} using the data from \cite{bernard16}. Note that many parts of the outer disc remain unexplored beyond Monoceros.}
\end{figure*}

We model the orbit of Sgr from the infall radius to the present-day for models of the Sgr progenitor with two different masses. We also adopt two different concentrations for our Sgr models, which results in two different characteristic central densities. The density profile determines how rapidly the satellite disrupts owing to tidal forces \citep{penarrubia10b} and hence has significant influence on the mass remaining bound to the satellite during final infall. This in turn affects the impact of Sgr in shaping the vertical structure of the disc, particularly at late times during the last passages through the disc plane. In the final part of this contribution we show how the LMC couples with  Sgr to sculpt the structure of the outer stellar disc. In section 2, we present the numerical set-up of our experiments. In section 3, we discuss the orbits of the different Sgr models and their impacts on the solar neighbourhood. In section 4, we compare the outcomes of each models on the structure of the MW disc in physical and observational space. In section 5, we use one of the simulations that is successful in simultaneously reproducing Monoceros/GASS, A13 and TriAnd-like features to explore their origin and contrast with earlier results. In section 6, we explore the response of the MW to both Sgr and the LMC and ask whether the LMC can still leave any visible signatures on the Galactic stellar disc. Our results and their implications are discussed in section 7 and we conclude in section 8.

\section{Numerical methods}

\subsection{Model for the MW}

For the MW, we adopt a model that is close to the one in \cite{gomez15} and \cite{mcmillan11}. We use the initial conditions generator {\sc galic} by \cite{yurin14} for which we have implemented a component taking into account adiabatic contraction as formulated by \cite{blumenthal86}. The virial mass of the Galaxy is $\sim10^{12} \rm{M_{\odot}}$ with a dark halo represented by a Hernquist sphere as in \citep{Springel2005c} with mass $M_{h}=10^{12}\,\rm{M_{\odot}}$ and scale length $a_{h}=52\, \rm{kpc}$\footnote{Prior to applying adiabatic contraction.}. The stellar disc is modeled as an exponential disc with scale radius $R_{d}=3.5 \,\rm{kpc}$ and scale height $h_{d}=0.53 \,\rm{kpc}$ and a total mass $M_{d}=6.0\times 10^{10} \rm{M_{\odot}}$. The bulge is represented by a Hernquist sphere with mass $M_{b}=1\times10^{10} \rm{M_{\odot}}$ and a scale radius of $a=0.7 \,\rm{kpc}$. For these choices of parameters the rotation curve peaks at a value of $\sim 239 \rm{km/s}$ as parametrised by \cite{mcmillan11}. We set $\sigma_{R}/\sigma_{z}=2$ and for our model, the Toomre Q parameter is above 1 everywhere in the disc. This is different fom our models in \cite{laporte16} because here, we would like to achieve discs that would be stable for timescales longer than 9 Gyr\footnote{The MW model in \cite{laporte16} becomes bar unstable after 3 Gyr. The new fiducial model presented here was run in isolation for more than 10 Gyr and showed no emergence of spiral arms or bar formation.}. The justification of this choice may seem arbitrary, but we do not wish to misinterpret our results due features such as spiral arms or a bar which could form suddenly during the course of the simulation\footnote{This choice is strictly numerical to ease the interpretation of the simulation, in reality both spontaneous spiral formation due to instabilities and merger induced perturbations may be studied jointly}. Given the large uncertainties on the structure of the MW and the cost of the simulations, we focus on just one fiducial model.

\subsection{Models of Sagittarius}

We consider two models for Sgr dSph of varying virial masses $M_{200}$\footnote{defined as the mass enclosed within a radius $r_{200}$ containing 200 times the critical density $\rho_{c}$ of the Universe} in the range $6.0\times 10^{10} \rm{M_{\odot}} - 1.0 \times 10^{11} {\rm{M_\odot}}$. These are parametrised as \cite{Hernquist1990} profiles with masses $M$ and scale radii $a$ chosen to closely match the corresponding NFW \citep{Navarro1996} halos but with steeper fall-off at large radii \citep{Jang2001}. Additionally, due to the scatter in the mass-concentration relation \citep{gao08,ludlow14}, we also consider two more progenitors that are twice as concentrated for the same given virial mass. These values are listed in Table 1. For each of the N-body models, we also include an extra self-gravitating component of a Hernquist sphere with mass $m_{*}=6.4\times10^{8}\,\rm{M_{\odot}}$ and scale radius $a_{*}=0.85\,\rm{kpc}$ to represent stars embedded within the dark matter halo \citep{ostholt12}. These were also generated with GalIC and the stability and disruption of these models were checked against equivalent models generated using distribution functions methods \citep{Kazantzidis2004, Laporte15} which led to similar results, confirming the large improvement from GalIC over previous methods \citep{Springel2005c}.

\begin{table}
 \centering
 \begin{minipage}{130mm}
  \begin{tabular}{@{}llrrrrlrlr@{}}
  \hline
Run & $ M_{200}$ & $c_{200}$ & $M_{h}$ & $a_{h}$ \\
 &   $10^{10}\rm{M_{\odot}}$ &  & $10^{10}\rm{M_{\odot}}$ & \rm{kpc} \\
  \hline
H1 & $10$  & $13.0$ &$14$& $ 13$ \\
H2 & $10$  & $26.0$ & $14$ & $ 7  $ \\
L1 & $6.0$ & $14$ & $8$&$ 16 $ \\
L2 & $6.0$ & $28 $ & $8$ &$ 8 $ \\
\hline
\end{tabular}
\end{minipage}
\caption[LMC ICs]{Structural parameters considered for the progenitor of the Sgr dSph and the corresponding Hernquist profile conversions following \cite{vdmarel12}}
\end{table}

\subsection{Simulations}

The N-body simulations presented here are run with the tree-code {\sc gadget-3} code \citep{Springel2005a}. We represent the Galaxy with particle masses of $m_{h}=2.6\times10^{4} \,\rm{M_{\odot}}$, $m_{d}=1.2\times10^{4} \,\rm{M_{\odot}}$, $m_{b}=1.0\times10^4\,\rm{M_{\odot}}$ for the dark matter, disc and bulge components respectively. We choose softening lengths of $\epsilon_{h}=60 \,\rm{pc}$, $\epsilon_{d}=\epsilon_{b}=30\,\rm{pc}$. This allows us to represent the Galactic disc with 5 million particle and the dark matter (DM) halo with 40 million particles such as to minimise the effect of heating between the different species. For Sagittarius, we use masses of $m_{DM_{sgr}}=2.0\times10^{4} \,\rm{M_{\odot}}$ and $m_{*_{sgr}}=4\times10^{3} \,\rm{M_{\odot}}$ for the dark matter and stellar particles respectively. We use softenings of $\epsilon_{DM_{sgr}}=60\,\rm{pc}$ and $\epsilon_{*_{sgr}}=40 \,\rm{pc}$. We choose an opening angle of $\theta=0.6$ degrees for the tree. An adaptive time-stepping scheme is used assigning time steps to individual particles following the condition $\Delta t_{i} =\tau/2^{j}<\sqrt{2\eta \epsilon_{i}/ \left|\mathbf{a_{i}}\right|}$, where $\eta=0.02$ is an accuracy parameter, $\mathbf{a_{i}}$ is the gravitational acceleration at the position of particle $i$ with softening length $\epsilon_{i}$ and $\tau$ is the base timestep which we choose it to be $\tau \le 10\, \rm{Myr}$.

\section{Finding `realistic' Sgr models}

While our goal is not to model Sgr or its resulting stellar stream in great detail, we do want to follow an object on a Sgr-like orbit with appropriate mass as close as possible to the real dwarf in order to study the physical role it must have played in structuring the disc. In this section, we use properties of the stream and remnant  to check the above. In addition, we want to follow the full interaction, hence we consider orbits starting around the virial radius $R_{200}$, unlike previous self-consistent models focused on the disc dynamics.

\subsection{Orbits from stream}

Our starting point to determine an orbit for Sgr from its infall radius is taken from the work of \citep{purcell11}. In short, these initial conditions assume that Sgr is already truncated at the instantaneous tidal radius and follow the orbit of the satellite by setting it at $80 \,\rm{kpc}$ from the Galactic center in plane of the disc with a velocity vector in the direction of the Northern Galactic Pole traveling at $v=80\,\rm{km/s}$. The model manages to reproduce the shape of the Sgr stream on the sky and its present-day location and line-of-sight velocity reasonably. Thus, we take these initial conditions for our starting point for our backwards integration which includes an implementation for dynamical friction \citep{Chandrasekhar1943} to determine the velocity and position vector of Sgr's progenitor around the MW's virial radius ($R_{200}=214 \,\rm{kpc}$). This is a similar procedure presented in \cite{kallivayalil13}, \cite{gomez15} or \cite{laporte16} to determine the initial conditions for the LMC in a first infall scenario. We then recalibrate the orbit with the aid of low- resolution N-body simulations and changing the Coulomb logarithm until we find a model that results in a present-day Sgr stream that is in reasonable agreement with the M-giants from \cite{majewski03}.
% This is the same procedure presented in {\bf \cite{kallivayalil13, gomez15, laporte16}} to determine the initial conditions for the LMC in a first infall scenario. We then recalibrate the orbit with the aid of low-resolution N-body simulations until some convergence is met by changing the Coulomb logarithm \citep{laporte16}.
\begin{figure}
\includegraphics[width=0.5\textwidth,trim=0mm 0mm 0mm 0mm,clip]{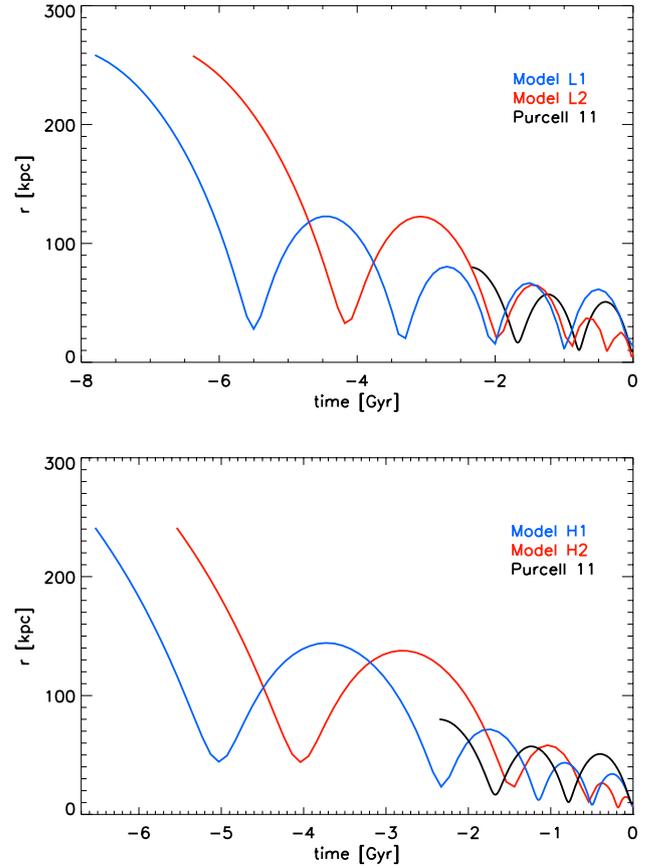}
\caption[]{Orbital decay as a function of time for all four models: L1, L2, H1 and H2. We also show the orbital decay from the experiments of Purcell et al. (2011) which have been typically used in all studies of the response of the disc to Sagittarius. We note that these models which consider similar mass models of the MW to us in fact missed not just one pericentric passage but two. As will be seen later, these have strong repercussions on the structure of the disc and its late time evolution. We see that under these models the MW may have interacted with the Sgr dSph within the last 6-8 Gyr corresponding to redshifts just below $z\sim1$.}
\end{figure}
Figure 2 shows N-body simulation orbital decay orbits for the different Sagittarius models. The more concentrated models have shorter orbital decay timescales because their higher density means that they start with and retain a higher bound mass and hence experience more efficient dynamical friction. This means that the total length of integration varies between 5.6 to 8 Gyr. The final shape of the stream on the sky and line-of-sight velocity of all four models are in reasonable agreement with observations \citep{majewski03}. This gives us confidence that the dynamical timescales and strength of influence on the Milky Way disk are in the right regime. This is discussed further in the Appendix A for the interested reader.

\subsection{Disc survivability}

Before embarking on further analysing our simulations, it is interesting to ask how discy the MW remains subject to its interactions in the different models considered here. Figure 3 shows the evolution of the circularity parameter that we defined as in \cite{scannapieco09} as $\epsilon_{v}=j_{z}/rv_{c}(r)$, where $j_{z}$ is the vertical component of star particle angular momentum vector and $v_{c}(r)$ is the circular velocity for model H2 which is the most extreme one of the four considered. We see that at the present-day, the galaxy remains discy (amounting to a decrease of 20 percent in number of particles with $\epsilon_{v}\ge0.7$ for model H2) with a small velocity dispersion dominated component, which is part of the bar that is formed in the simulations during the last pericentric passage of Sgr.

\begin{figure}
\includegraphics[width=0.5\textwidth,trim=0mm 0mm 0mm 0mm,clip]{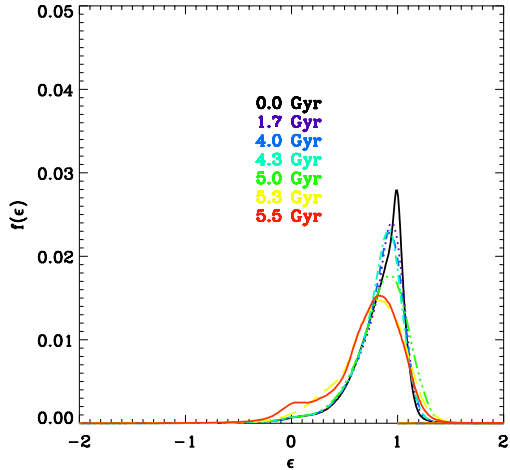}
\caption[]{Evolution of the circularity parameter $\epsilon_{v}=j_{z}/rv_{c}(r)$, as defined by Scannapieco et al. 2009. The Sgr's interaction with the disc causes a tail of velocity dispersion supported stars originating from the disc which belong to the bar, but most stars remain still on circular orbits.}
\end{figure}
\subsection{The solar neighbourhood}

Although in our models we note the existence of strong oscillations in the structure of the outer-disc, the viability of any of our models should be put to test by first looking at the observational constraints on the vertical structure of the disc in the solar neighbourhood. These include for example, the measurement of the mean vertical motion of stars near the Sun \citep{widrow12, williams13,carlin13} and the residual number density profile of the disc \citep{widrow12,ferguson17}. For the analysis, we position ourselves around the Sun ($(x,y)=(-8, 0)$ kpc) and select particles within a cylinder of radius $R=1\,\rm{kpc}$ and calculate the mean vertical velocity of stars as a function of height about the midplane. This is shown in Figure 4 for all four models which we compare to data from \cite{widrow12}. 
%Because, the streaming motions around solar neighbourhood-like regions are quite sensitive to the time between snapshots $\Delta t\sim10 \,\rm{Myr}$ (due to the passage of a spiral arm), we select the timesteps which give the most reasonable matches (within $\Delta t \sim 30 Myr$) from where Sgr reaches its present-day location. 
These authors found that the vertical streaming motions in the solar neighbourhood varied in the range of $-10\, \leq V_{z}/\rm{km/s}\leq10 $. We find similar ranges in the simulations. Quantitatively, not all simulations result in a good match (this is not expected as this is a level of detail that our experiment cannot adequately address), but their behaviour is similar to that of the data, except perhaps for model H1 which seems to be dominated by a bending wave. On the other hand, we also note that model H2 shows agreement with the observed data, within the errors.

\begin{figure}
\includegraphics[width=0.5\textwidth,trim=0mm 0mm 0mm 0mm,clip]{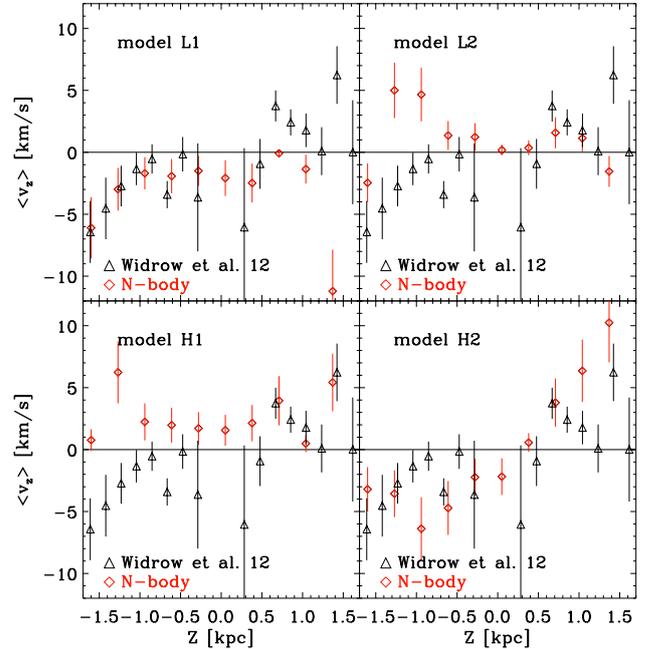}
\caption[]{Vertical streaming motion around cylinder of $\sim1 \,\rm{kpc}$ in radius centered at the Sun. Our models show streaming motions that vary between -10 km/s and 10 km/s as seen in the observations as well as oscillatory motion. The velocities we find do not violate constraints in the solar neighbourhood.}
\end{figure}

Next, we also calculate the number density $n(Z)$ of stars as a function of height $Z$ in a similar fashion as in \cite{gomez13} and \cite{laporte16}. In Figure 5, we present the residual of the number density $\Delta n(Z)$ as a function of height about the midplane. This is calculated by substracting $n(Z)$ by the azimuthally averaged value $<n(Z)>$ measured across a cylindrical annulus of width $\Delta R= 1\,\rm{kpc}$ at $R=8\,\rm{kpc}$ from the Galactic Center and averaging, giving $\Delta n = (n(Z)-<n(Z)>)/<n(Z)>$. We see that the amplitude of the fluctuations correspond to the magnitude of the oscillations seen in the observations rather well. However, the phases are not always correct. Nonetheless we note that some models show rather good agreement: this is the case of the L2 and to a certain degree the H2 models. This was not the case with previous simulations of the interaction of Sgr dSph with the MW \citep{purcell11,gomez13,laporte16}. We thus conclude that none of our models violate structural and kinematical constraints in the solar neighbourhood.

%In these comparisons, it is important to note that we have not tried to find a perfect match model to the Sgr orbit with its present-day position and velocity. Given the uncertainties in the mass of the Milky Way, and the dynamical structure of the disc, this is not yet feasible, especially to model the structures we see in the solar neighbourhood with N-body simulations. However, given that the amplitude and in some instances the phases of the signal we see are already qualitatively very similar to the observations and within the large error bars this is rather promising. Given that the H2 model semi-quantitatively reproduces Monoceros/GASS and A13/TriAnd-like features (in terms of extent on the sky but also distances and phases to the Sun), we can use this realisation as a case study to understand the origin of these features further and where possible identify the differences in this model compared to previous ones \citep{purcell11, gomez12, price-whelan15, laporte16}. This is the subject of the next section.

\begin{figure}
\includegraphics[width=0.5\textwidth,trim=0mm 0mm 0mm 0mm,clip]{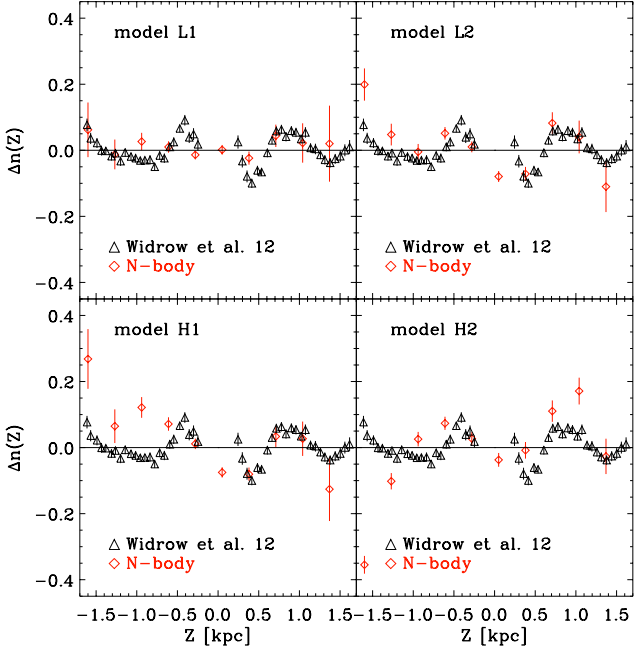}
\caption[]{Number density residual around a cylinder of $\sim1 \,\rm{kpc}$ in radius about the Sun. Similarly to the observations, our models also show signs of oscillations across the plane of the disc. The amplitudes of the oscillations agree between the models and observations within the error bars margin (except at some edges), however the phases do not always correspond. However we note that model H2 and L2 show rather good agreement in amplitude and phases.}
\end{figure}

\section{Present-day structure and kinematics of the Galactic disc induced by the Sgr dSph}

Having built confidence that we are in the right regime, we compare the present-day vertical structure of the Milky Way disc from the different realisations (H1,H2,L1,L2) of the interaction of Sgr with the disc. In order to analyse the structure of the disc, we first re-center the halo using a shrinking sphere algorithm \citep{Power2003}. We then align the galaxy onto the plane perpendicular to the disc's angular momentum vector, through an iterative process as in \cite{gomez13} and \cite{laporte16}. We do this by selecting all disc particles within a cylinder radius of $8 \, \rm{kpc}$ and rotate the system such that the $L_{z}$ component of the angular momentum vector is aligned with the $z$-axis. We then decrease the height of the cylinder and repeat the procedure until convergence is met. The goal of this section is two-fold: we want to illustrate the predicted structures across the whole disc in terms of mean velocity and mean vertical height of the disc and compare where possible (as the stellar disc has not been observed in its entirety) to observations of the outer disc. In section 4.1, we discuss the outcomes of the different simulations in terms of the predicted mean vertical structure and motions of the discs. In section 4.2 we compare the simulations to known observations of the disc towards the Galactic Anticentre at different Galactocentric radii which were presented in the introduction (see Figure 1).

\subsection{Vertical structure and kinematics}

\begin{figure}
\includegraphics[width=0.5\textwidth,trim=0mm 0mm 0mm 0mm,clip]{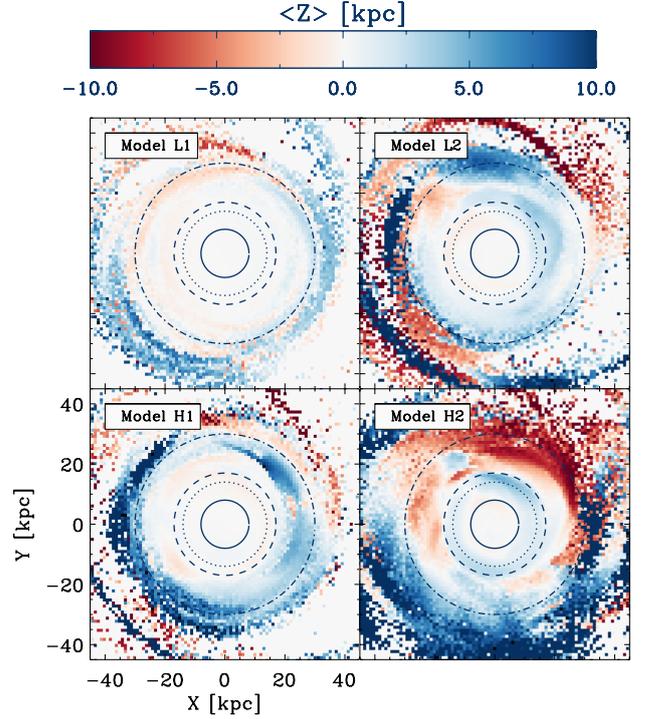}
\caption[]{Mean vertical height map of disc stars about the midplane. The radii of Monoceros North, South and TriAnd are denoted by the dotted, dashed and dash-dotted lines at $R=16,18,30 \,\rm{kpc}$. We note that all models exhibit stars at extreme locations above and below the plane at large radii $R\ge 30\,\rm{kpc}$, above $\sim10$ scale heights. Because of the multiple pericentric passages of Sgr the vertical oscillations through their constructive and destructive interference show less of ring-like appearances but more arc-like features. We note that there is a degeneracy between models of different progenitor mass with the central density with the L2 model closely resembling the H1 model in terms of amplitude fluctuations in mean height and wavelength. Particularly, we note that model H2 reproduces a Monoceros Ring-like feature which coincides well with the locations where it it is observed in the Galaxy (within a $\sim1\, \rm{kpc}$ margin).}
\end{figure}

Figure 6 shows maps of the mean height of disc stars about the midplane, calculated inside square pixels of $1 \,\rm{kpc}$ in size. In this coordinate system, the Sun lies at $x,y)=(-8,0)$. We have also overplotted concentric rings of radii=$8,16,18,30$ which coincide with the positions of the Sun, the Monoceros Ring in the North and South as determined by the Pan-STARRS study of \cite{morganson16} and the TriAnd clouds \citep{sheffield14,price-whelan15}. Although all four models reasonably reproduce the Sgr stream on the sky (see Appendix A), they each predict different structure within the disc with material being kicked to different distances. This suggests that features such as Monoceros/GASS, A13 and TriAnd may provide extra constraints to discriminate between different models for the MW/Sgr interaction. 

%{\bf change this to something more physical obs later!!!} We note several structures in these models in particlar, a clear Monoceros Ring feature present for model H2, with a close ring appearance. It also shows a North South asymmetry that coincides well  - within $\sim1 \,\rm{kpc}$ - with the observed positions of the Monoceros Ring in the North and South in observations of \cite{morganson16}. The amplitude in vertical mean height of stars can vary between $-5 \leq Z/\rm{kpc} \leq 5$. Furthermore, out to radii of $R\~30\,\rm{kpc}$ disc material in the direction of TriAnd can reaches heights of $Z\sim -10 \,\rm{kpc}$ in good agreement with observations \citep{martin07, sheffield14}.

Within a Galactocentric radius of $R\sim 16 \,\rm{kpc}$, from a model to another, the amplitude in vertical mean height of stars can vary between $-5 \leq Z/\rm{kpc} \leq 5$ and just below $|Z|\sim1\,\rm{kpc}$. Furthermore, out to radii of $R\sim30\,\rm{kpc}$ disc material can be excited to absolute heights of $|Z|\sim5\,\rm{kpc}$ to $|Z|\sim10\,\rm{kpc}$ about the midplane. We also see that there is a clear degeneracy in the amplitudes of the oscillations of the disc between intitial mass and central density of the progenitor for Sgr.  This is notably the case for model H1 and L2, which produce vertical asymmetries of rather similar amplitudes. This is tied to the similar final bound mass that these models produce for the Sgr remnant. The recorded present-day Sgr masses within 2.5 kpc of the remnant are $\sim4, 2, 3, \,\rm{and}\, 0.4\times10^{9} \,\rm{M_{\odot}}$ for models H2, H1, L2 and L1 respectively.

Figure 7 shows maps of the mean vertical velocity of disc stars about the midplane, calculated in the same fashion as in Figure 6. We note that the fluctuations can vary between practically  almost no motion in the inner galaxy to 10 km/s. In the outer-disc, at radii $R\ge10$, these fluctuations vary from $|V_{z}|\sim10\,\rm{km/s}$ to as much as $V_{z}\sim50\,\rm{km/s}$. Not surprisingly the same degeneracy noted between models H1 and L2 is also being translated in velocity space. 

\begin{figure}
\includegraphics[width=0.5\textwidth,trim=0mm 0mm 0mm 0mm,clip]{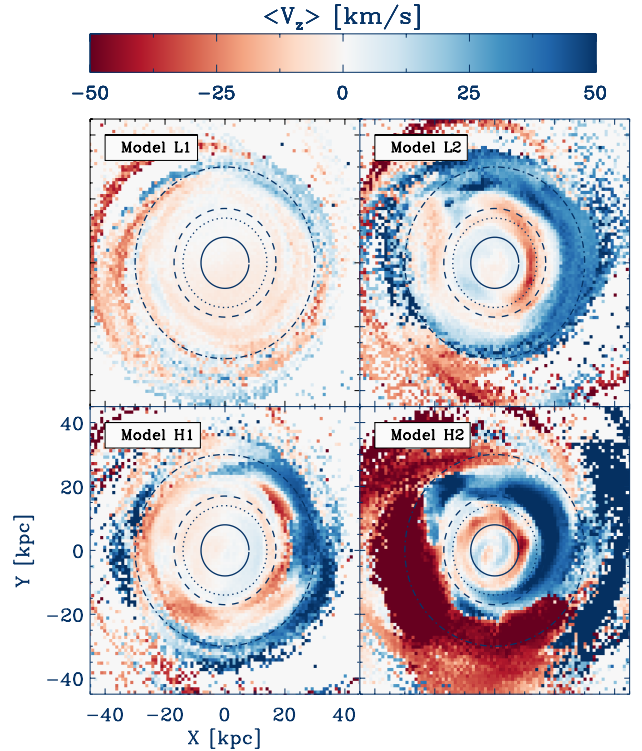}
\caption[]{2D vertical velocity maps of the Milky Way disc for models L1, L2, H1 and H2 respectively. The rings are the same as in Figure 6. The bulk vertical motions of the disc can vary from a few tens of $\rm{km/s}$ within the solar neighbourhood to $|V_{z}|\sim50\,\rm{km/s}$ at $R\geq16\,\rm{kpc}$.}
\end{figure}

Figure 8 and 9 show the mean height and velocities of stars between the different models as a function of Galactocentric radius for different longitude ranges through wedges spanning an opening angle of 10 degrees in the directions $l\sim 0^{\circ}, 90^{\circ}, 180^{\circ}, 270^{\circ}$. We note that perturbations that are diametrically opposed do not always mirror each other. Indeed, along $l\sim90^{\circ}$ and $l\sim 270^{\circ}$ the behaviours are not symmetric and can be opposite (right panel model L1 Figure 8) or completely independent (model H2 right panel Figure 9). This calls for some caution when interpreting observations of, for example, the Monoceros Ring as symmetric ``ring'' structures especially when trying to calculate its total mass \citep{morganson16}. This also goes for interpreting features such as A13, TriAnd as part of a clear symmetric oscillatory feature of the disc. Because Sgr has gone around the disc several times in its past, multiple vertical perturbations get excited and interfere with one another such that the simple picture of symmetric ripples in the disc fails to describe the current state of the Galaxy. 

We note that although degeneracies in amplitude exist, characterising the wavelength of the response of the disc  along different lines of sight may be an additional discriminant to constrain one impact model over another. It is evident that more detailed studies of the outer disc are necessary along different lines of sight in order to understand its structure. Our models clearly do not predict rings but complex oscillatory features that vary with azimuth in wavelength as a result of the multiple interactions of Sgr with the disc. We also find that Sgr also contributes to gradually flaring the disc (this is discussed in section 5.1)

\begin{figure}
\includegraphics[width=0.5\textwidth,trim=0mm 0mm 0mm 0mm,clip]{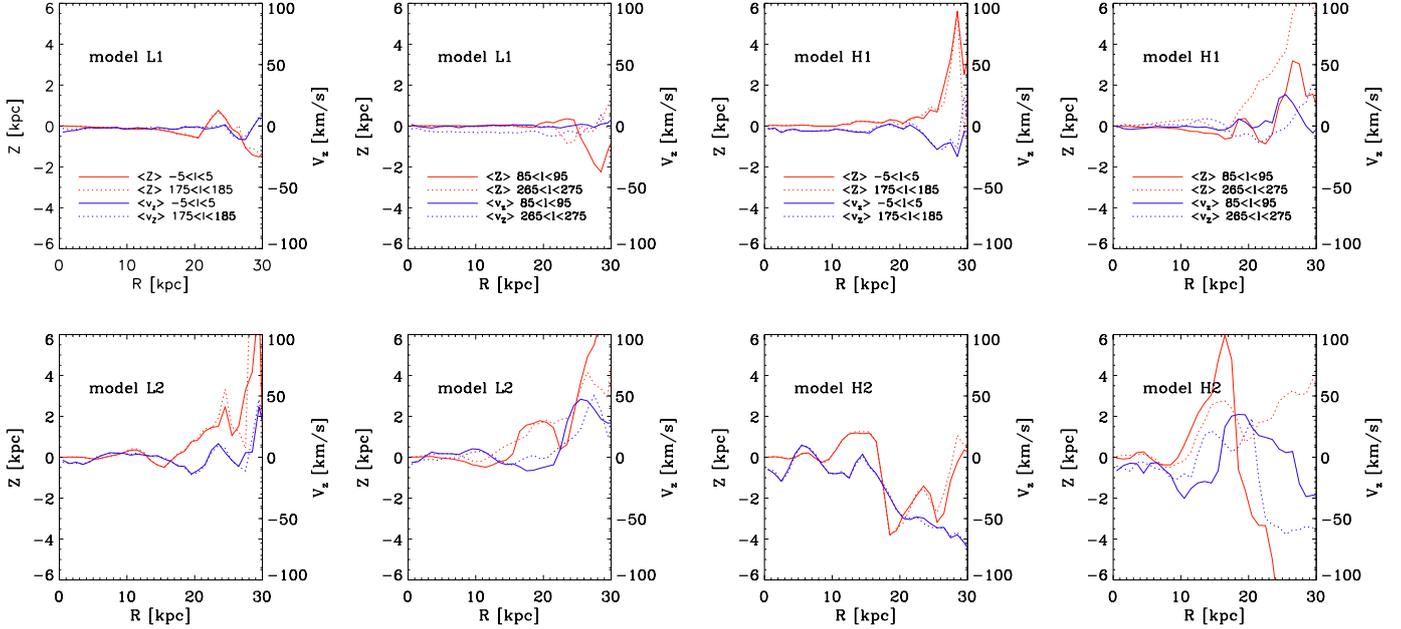}
\caption[]{Mean heights and velocities (shown in blue and red respectively) of stars as a function of Galactocentric radius along different directions in longitude for models L1 and L2. Dashed lines correspond for to mean heights and velocities on diametrically opposite directions.}
\end{figure}

\begin{figure}
\includegraphics[width=0.5\textwidth,trim=0mm 0mm 0mm 0mm,clip]{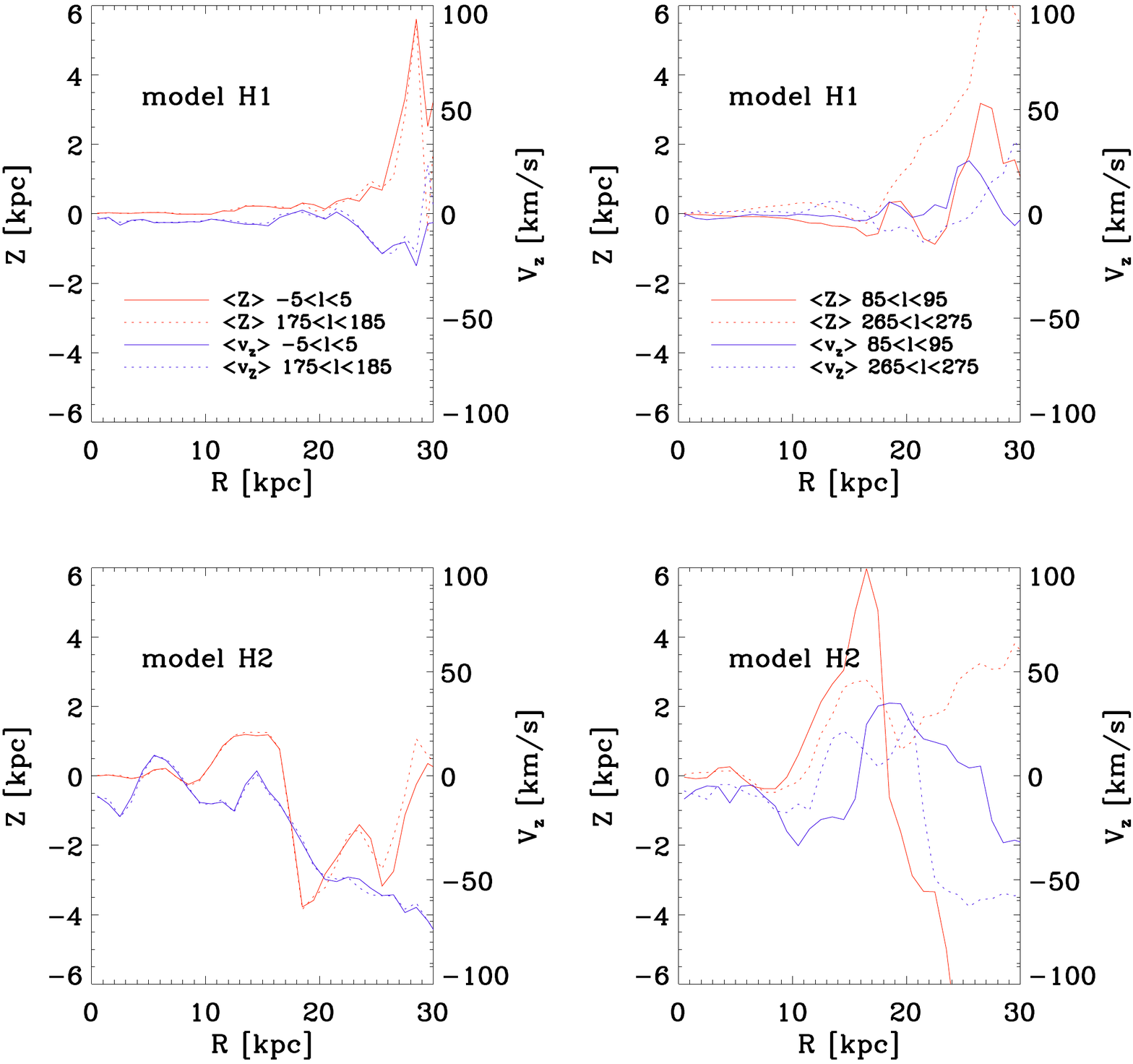}
\caption[]{Same as Figure 8 for models H1 and H2.}
\end{figure}

\subsection{Observational perspective}

In this section, we inspect the structure of the disc from an observational perspective. Monoceros has recently been mapped to unprecedented detail in the Pan-STARRS data \citep{slater14,morganson16}, showing that its extent lies above and below $|l|\sim30^{\circ}$ of latitude. Figures 10, 11 and 12 show maps of N-body counts in the Galactic coordinates $(l,b)$ (for which we bin our data in bins of $2^{\circ}$ on a side) split in slices of heliocentric distances, roughly corresponding to the radii where Monoceros/GASS and A13/TriAnd are observed ($5<h_{d}/\rm{kpc}<10$, $10<h_{d}/\rm{kpc}<13$ and $13<h_{d}/\rm{kpc}<25$  for Monoceros \citep{morganson16}, GASS \citep{crane03} and for A13/TriAnd \citep{li17,sheffield14} respectively).  

From Figure 10, we see that different models predict different scales for disc material being kicked up to high latitudes. Only the more concentrated models are able to produce features reminescent of the Monoceros Ring in terms of its extent on the sky \citep{slater14}. Model H2 in particular, is capable of bringing disc material out to $|b|\sim30^{\circ}$ as observed for the Monoceros Ring \citep{slater14}, while models L2 and H1 just marginally do so. On the other hand, Model L1 is not able to to excite disc material to the observed high latitudes of the Monoceros Ring. In fact, in the 2D mean height map of model H2, the Monoceros Ring is mostly bound by two rings at $R=16\,\rm{kpc}$ and $R=18\,\rm{kpc}$ which coincides very well with the northern and southern parts of Monoceros \citep{morganson16}.

In Figure 11, we show the distribution of the N-body star particles at similar heliocentric distances to the GASS overdensity stars of \cite{crane03}, which are overplotted as fuschia open squares. Again, we see that the low-mass Sgr model L1 struggles to kick disc material out to the latitudes where GASS stars lie. The more concentrated model L2 or massive Sgr models ($M\sim10^{11}\,\rm{M_{\odot}}$)  perform better with the H2 model again being mostly favoured by the data. While individual models may reproduce certain features in a given radial extent of the disc, it is not a sufficient condition to guarantee their success overall, illustrating the importance of considering the Galactic disc {\it globally} and not just locally. Indeed, while model L2 manages to kick-up disc material in the region where TriAnd is observed, it barely manages to reproduce a Monoceros-like feature. Thus it is important to assess how the various models perform at heliocentric distances beyond $10\,\rm{kpc}$, where disc material has been identified through stellar populations studies \citep{price-whelan15} namely for TriAnd and also A13 (Bergemann et al. submitted). 

In Figure 12, we present  our N-body stars in $(l,b)$ at distance cuts beyond beyond GASS, for which we overlay the stars identified to belong to TriAnd and A13\citep{sheffield14,li17}. Again, we see that model L1 consistently fails at reproducing any observed features of the outer disc. On the other hand models L2, H1 perform marginally well. This leaves model H2 as the only one able to simultaneously excite all four features associated with the outer-disc towards the Anticenter. We shall see in section 5 that it is also able to account for the location of the Cepheids seen in the direction of the Galactic center located at $R\sim15-20\,\rm{kpc}$ with absolute heights of $|Z|\sim2\,\rm{kpc}$.

This is not to say that the Sgr progenitor must have had the same properties as our H2 model progenitor, but rather hints towards the fact that the mass of the Sgr dwarf spheroidal progenitor must have been around or above $M_{halo}\sim6\times10^{10}\, \rm{M_{\odot}}$ and that the main body of Sgr dSph must have impacted the stellar disc of the MW. This can be understood because the tidal effects of Sgr become increasingly important - if we want Sgr to be solely responsible for these features - for the evolution of the outer-disc during the last pericenters, as we shall see section in 5.3. Model H2 is particularly interesting because it demonstrates that {\it all the observed features towards the Anticenter can understood as originating from the ongoing disruption of a single satellite} as opposed to separate events of satellite disruptions \citep{penarrubia10,sheffield14} along the midplane or tides from multiple encounters \citep{kazantzidis08}. Models with masses for Sgr dSph progenitor with ($M_{200}\leq6\times10^{10}\,\rm{M_{\odot}}$) at face value seem to be ruled out, lending support for a more massive Sgr dSph as has been suggested in the past \citep{jiang00} and more recently in \citep{gibbons16} who use the velocity dispersion of the Sgr stream as a discriminator.

Recently, \cite{deboer17} have measured proper motions using a combination of the Gaia DR1 release and the SDSS. They measure gradients in the proper motion of the stars located at a heliocentric distance of $10\,\rm{kpc}$ in a region of $160\leq l\leq230$ and $15\leq b\leq40$ in the range of $-1.5\leq\mu_{b}/(\rm{mas/yr})\leq1.5$ and $-5\leq\mu_{l}/(\rm{mas/yr})\leq1$. We have transformed our model velocities into the the ($\mu_{l}, \mu_{b}$) space and find that our model H2 agrees qualitatively. We transform our positions and velocity vectors into a heliocentric coordinate system and take into account the solar motion to calculate $(l,b,d, v_{rad}, \mu_{l}, \mu_{b})$. In Figure 13, we present the maps of the mean proper motions for Monoceros in $(l,b)$. Our simulations match qualitatively best the data from \cite{deboer17} at a distance of $d\sim10\,\rm{kpc}$ which is within the range of the distances derived for the Monoceros Ring \citep{ibata03,slater14,morganson16}. Across the disc, we note that the vertical motions vary in the range  $-2\leq\mu_{b}/(\rm{mas/yr})\leq-2$.  In the region where \cite{deboer17} observed Monoceros, this is varies within $-1.5 \leq \mu_{b}/(\rm{mas/yr}) \leq 1.5$, in agreement with the observations. We do not get the same gradient, although one may argue that the vertical motion feature at $(l,b)=(240,40)$ extends into the region around $l\sim170^{\circ}$ making it appear like we are reproducing the gradient that is seen. We note that this is rather tentative and given the relatively low particle numbers in this region further confirmation will need to wait for higher resolution disc models.  Although our models are of quite high resolution ($\sim 5 \times 10^{6}$ disc particles), measuring proper motions within a certain distance cut over a few $\rm{kpc}$ across the whole sky becomes subject to poorer resolution towards the Anticenter with particle numbers varying between 100 and of the order of a few 10s in bins of two degrees on a side\footnote{This is due to the exponential form of the disc profile and the fact that observational plots are positioned at the location of the Sun, thus one always probes larger Galactocentric distances towards the}. \cite{deboer17} also looked at the transverse motions in Monoceros and found an increase with latitude and also a decrease with longitude. Our proper motion maps show the same behaviour qualitatively as we move in longitude from $l\sim220^{\circ}$ to $l\sim150^{\circ}$ from $4.5\,\rm{mas/yr}$ to $3 \,\rm{mas/yr}$. A vertical gradient in the Anticentre region is also clearly visible.

Although we note that the portion of sky that \cite{deboer17} cover is relatively small compared to the full extent of Monoceros as seen in the Pan-STARRS survey data \citep{slater14} and that in some regions they may be dominated by contaminants, their derived maps and observations are in good qualitative agreement with our H2 model. A better match to the observations could possibly be found by exploring more mass models and with different kinematics for the disc, however our goal in this contribution was not to match the observations but to get a better understanding of the possible outcomes on the disc structure for different realistic Sgr dSph mass models and orbits. We will explore further the connection between observations and physical interpretation of kicked-up disc stars in the MW in a separate contribution (Laporte et al. in prep).% and also present higher resolution disc models ($N_{disc}\sim50 \times 10^6$ particles). 

\begin{figure}
\includegraphics[width=0.5\textwidth,trim=0mm 0mm 0mm 0mm,clip]{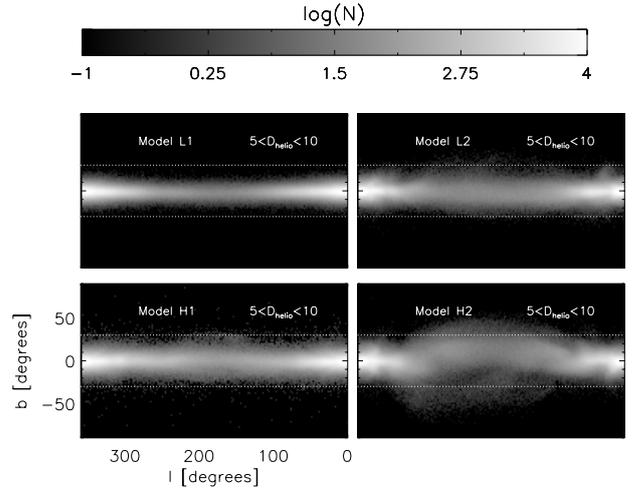}
\caption[]{Distribution of stellar N-body particles in $(l,b)$ within the distance range for Monoceros "North" and "South" as defined by Morganson et al. (2016) for all four models L1, L2, H1 and H2. The dashed white lines delineate the $|b|=30^{\circ}$ latitude location. The L1 model is not able to reproduce the distribution of stars above 30 degrees of latitude that extend in the data from Slater et al. (2014). Models L2 and H1 marginally kick disc material beyond $|b|\sim30^{\circ}$. Model H2 reproduces the full extent of Monoceros. Due to the large distance cut, feather-like features as in Slater et al. (2014) get washed away.}
\end{figure}

\begin{figure}
\includegraphics[width=0.5\textwidth,trim=0mm 0mm 0mm 0mm,clip]{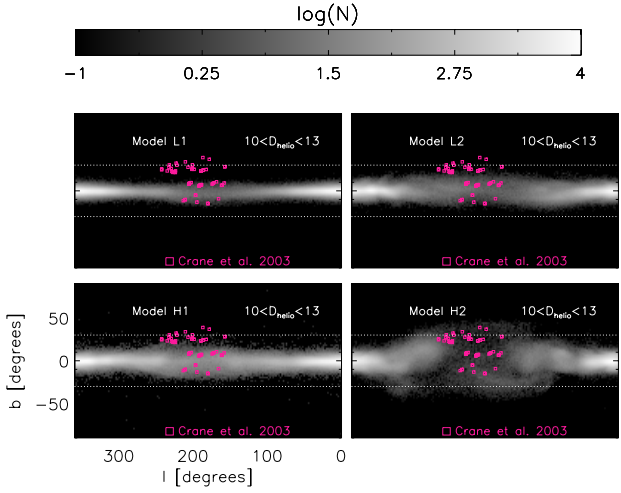}
\caption[]{Distribution of stellar N-body particles in $(l,b)$ within the distance range for GASS for all four models L1, L2, H1 and H2. The pink square symbols represent stars with $10\leq h_{d} \leq13$ from Crane et al. 2003 associated with GASS/Monoceros.  Model L1 is clearly ruled out. Note how the mass of the progenitor is not necessarily the defining factor in setting the final distribution of stars. Model L2 ($M_{h}\sim6\times10^{10} \rm{M_{\odot}}$) performs just as well as model H1. Only model H2 is able to kick disc material within the full range of latitudes observed for GASS stars.}
\end{figure}

\begin{figure}
\includegraphics[width=0.5\textwidth,trim=0mm 0mm 0mm 0mm,clip]{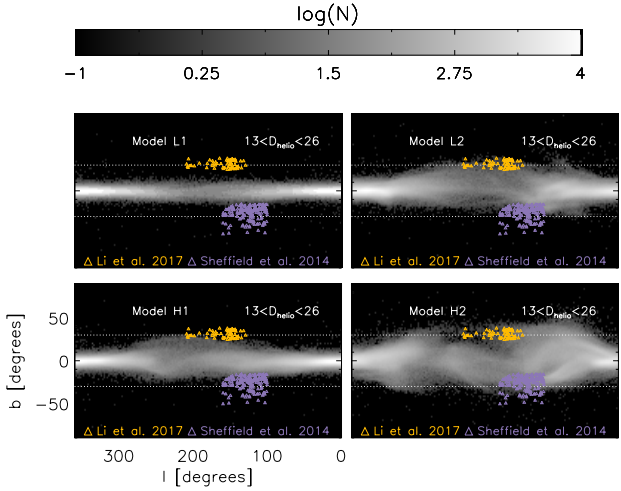}
\caption[]{Distribution of stars N-body particles in $(l,b)$ within the distance range where TriAnd and A13 are located for all four models L1, L2, H1 and H2. The orange triangles are the stars associated with A13 from Li et al. (2017) and the purple ones with TriAnd from Sheffield et al. (2014). Model L1 is not able to reproduce these features towards the Anticentre, a much higher central density progenitor is needed (model L2) in order to reproduce A13, though this model does reproduce the full extent of TriAnd. We have not tried to rotate the Galaxy but perhaps better modeling could find a better agreement.  The distribution of stars in model H2 covers the full extent of both TriAnd and A13.}
\end{figure}

\begin{figure}
\includegraphics[width=0.5\textwidth,trim=0mm 0mm 0mm 0mm,clip]{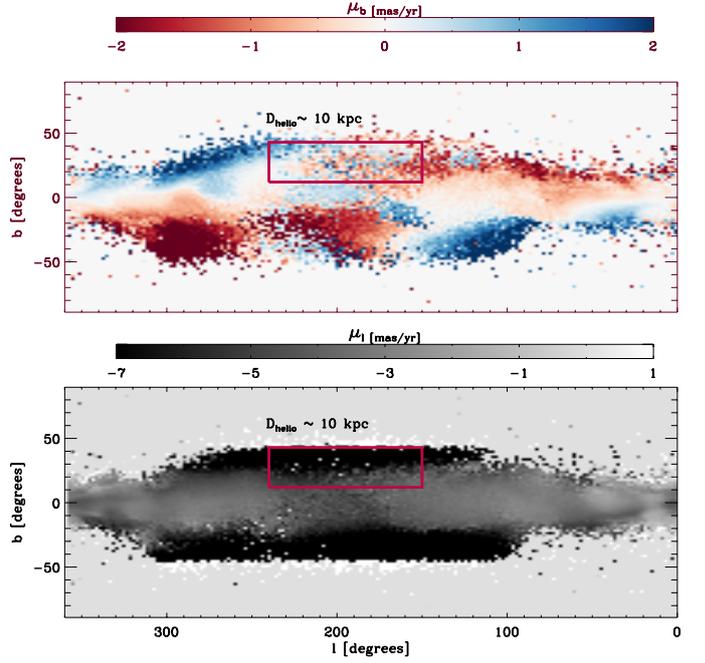}
\caption[]{Proper motions in the disc at a heliocentric distance of $h_{d}\sim10\,\rm{kpc}$. The magenta box marks the area observed by de Boer et al. (2017). {\it Upper Panel}: Mean vertical proper motion $\mu_{b}$ in $(l,b)$ across the disc. Vertical motions are clearly visible across the whole disc with a breathing pattern visble. The proper motions vary between $-2 \leq \mu_{b}/(\rm{mas/yr}) \leq 2$. Towards the region of Monoceros observed by de Boer et al. (2017), proper motions vary between $-1.5 \leq \mu_{b}/(\rm{mas/yr}) \leq 1.5$, and a qualitatively similar gradient trend is visible as in the observations. Quantitative comparison will have to await higher resolution simulations. {\it Lower Panel}: Mean horizontal proper motions $\mu_{l}$ in $(l,b)$ across the disc. Similarly to de Boer et al. (2017) we observe an increase in the azimuthal motion of stars with increasing latitude as well as a week drop as a function of longitude towards from $l\sim220^{\circ}$ to $l\sim150^{\circ}$.}
\end{figure}

\section{A Case Study: Model H2 and the emergence of Monoceros and TriAnd}

It is important to note that {\it all} previous models of the Sgr impact with the Milky Way, have consistently struggled quantitatively to explain the existence of disc material at low-latitudes \citep{purcell11, gomez13,laporte16}. While the experiment of \cite{kazantzidis08} could reproduce some qualitatively similar features, their models were not tailored to the Milky Way, but just considered a succession of mergers of satellites from a random cosmological DM-only simulation tree. However, it is not clear whether their realisation would reflect the typically quiescent merger history expected for the MW \citep{ruchti15} and would be consistent with the fact that the dominant contribution of the stellar halo remains Sgr \citep{bell08}. Our models here are able for the first time to reproduce the positions of different features separated by more than $10 \,\rm{kpc}$ and predict streaming motions that are in agreement with available data and relate them all to the accretion of the Sgr dSph. This is particularly true of model H2, which we examine more closely in this section. We begin by examining the structural evolution of the disc with time in section 5.1. In section 5.2 we follow the growth of the MW halo's dark matter halo wake. In section 5.3 we identify the mechanisms that sculpt the present-day structure of the outer disc of the MW by following in time the strength of the torques from the the inner DM halo and that of Sgr's DM halo in the plane of the disc.

\subsection{Structural Evolution of the disc}

\begin{figure}
\includegraphics[width=0.5\textwidth,trim=0mm 0mm 0mm 0mm,clip]{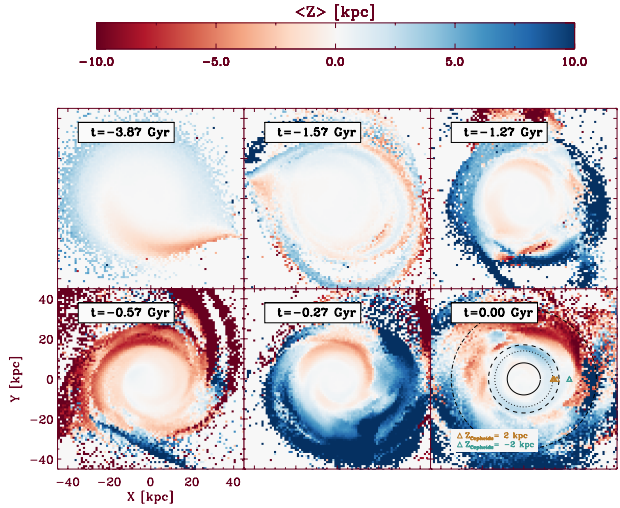}
\caption[]{ Evolution of the mean vertical height of the disc for model H2 at six different snapshots coinciding with each time Sgr crosses the z=0 plane. We show the location of the Cepheids from Feast et al. (2014) in the last panel corresponding to the present-day structure of the disc. During first pericentric passage, a clear m=1 warp is excited in the outer disc. Subsequent passages excite mean disc material to increasingly higher heights and the oscillations progress inwards to the center of the galaxy. The final galaxy shows complex morphology with several ring patterns which constructively and destructively interfere. Within this model, structures such as TriAnd and A13 could  come from material which existed already 6 Gyr ago. At the location of the Cepheids from Feast et al. (2014) at (x,y)=(15-20,0), the mean height varies from $2\,\rm{kpc}$ above the plane to $-2 \,\rm{kpc}$ below the plane. The only Cepheid that is observed at $R\sim20\,\rm{kpc}$ in their sample is situated at $|z|\sim2\,\rm{kpc}$ below the midplane, the others are situated above it at $z\sim 2\rm{kpc}$. This which coincides well with the vertical structure of model H2. }
\end{figure}

\begin{figure}
\includegraphics[width=0.5\textwidth,trim=0mm 0mm 0mm 0mm,clip]{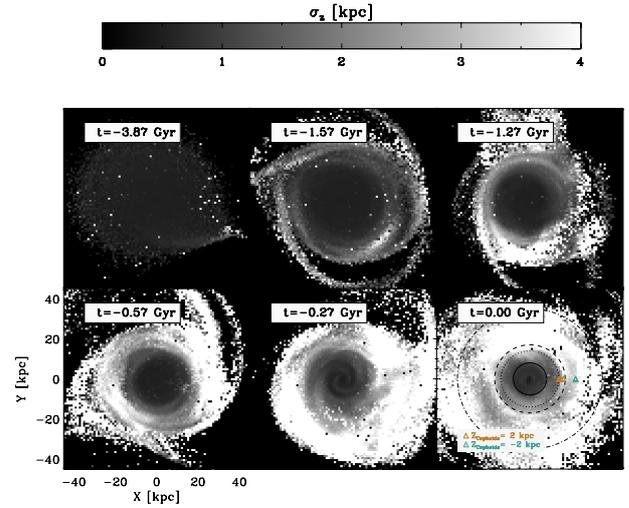}
\caption[]{Evolution of the dispersion in height of stars in the disc for Model H2. With subsequent passages, Sgr gradually flares the disc outside in with dispersions in heights of up to $\sigma_{z}\sim4\,\rm{kpc}$. Note that the inner regions remain much less affected. At the location of the Cepheids from Feast et al. (2014) at (x,y)=(15-20,0), $2\leq\sigma_{z}\leq4$ thus these stars are well accounted for in this model.}
\end{figure}

\begin{figure}
\includegraphics[width=0.5\textwidth,trim=0mm 0mm 0mm 0mm,clip]{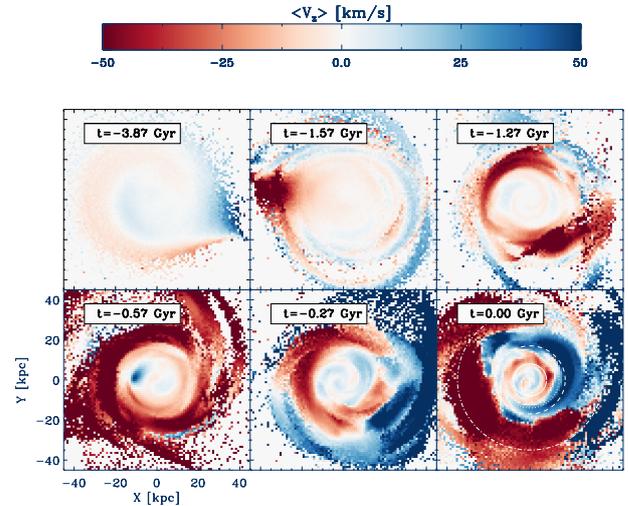}
\caption[]{Evolution of the mean vertical streaming velocity of the disc for model H2. As Sgr makes subsequent disc passages the vertical streaming motions gradually grow to reach values between $-50\,\rm{km/s}$ and $50\,\rm{km/s}$ beyond the Monoceros Ring and $-10\,\rm{km/s}$ and $10\,\rm{km/s}$ within the solar neighbourhood.}
\end{figure}

We begin this section by presenting a time series of the mean height of stellar particles about the midplane to illustrate how the vertical oscillations of the disc come to existence. This is shown in Figure 14 with a sequence of snapshots taken at instances after Sgr crossed the $Z=0$ plane. We notice that after the first pericentric passage ($r_{peri}\sim50\,\rm{kpc}$) the disc is already warped, which is characterised by a perturbation in the form of an $m=1$ distortion which develops already by $\sim1$ Gyr since time of infall. With subsequent passages, these vertical fluctuations amplify to the point that they can bring disc material to absolute heights of $|Z|\sim10 \,\rm{kpc}$ by the present-day at $R\sim30\,\rm{kpc}$. In passing, we also show the location of the Cepheids from \cite{feast14}, which are consistent with our simulation, suggesting that their extreme location may not exclude a vertical oscillation origin of the disc. In Figure 15 we show the dispersion in heights of the star particles about the midplane. We see that as the interaction with Sgr proceeds, the disc becomes gradually flared in an outside-in fashion. Thus, Sgr seeds vertical oscillations in the disc that ripple but overall the disc also flares as a result of the interaction.\footnote{In fact, the level of flaring between radii of $R\sim15\,\rm{kpc}$ and $R\sim20\,\rm{kpc}$ vary between $\sigma_{z}\sim2\,\rm{kpc}$ and $\sigma_{z}\sim4\,\rm{kpc}$, consistent with \cite{feast14}.}

Similarly, in Figure 16, we present the mean vertical velocity of the disc. We see that early on, the  outer disc already gets excited all the way into $R\sim 16 \,\rm{kpc}$ with streaming motions of the order of $10\,\rm{km/s}$ which subsequently get amplified during subsequent passages with velocities as large as $50\,\rm{km/s}$. While this is not directly comparable to the result of \cite{deboer17} because these authors measure the velocity of Monoceros as a function of latitude, our average velocity for the region where Monoceros is in line with the amplitudes of the velocities measured in the observations. Although multiple interactions create interferences in the vertical patterns, we are still able to pick up anticorrelations between $\langle Z \rangle$ and $\langle V_{z} \rangle$ which remain visible, modulo some exceptions.

\begin{figure}
\includegraphics[width=0.5\textwidth,trim=0mm 0mm 0mm 0mm,clip]{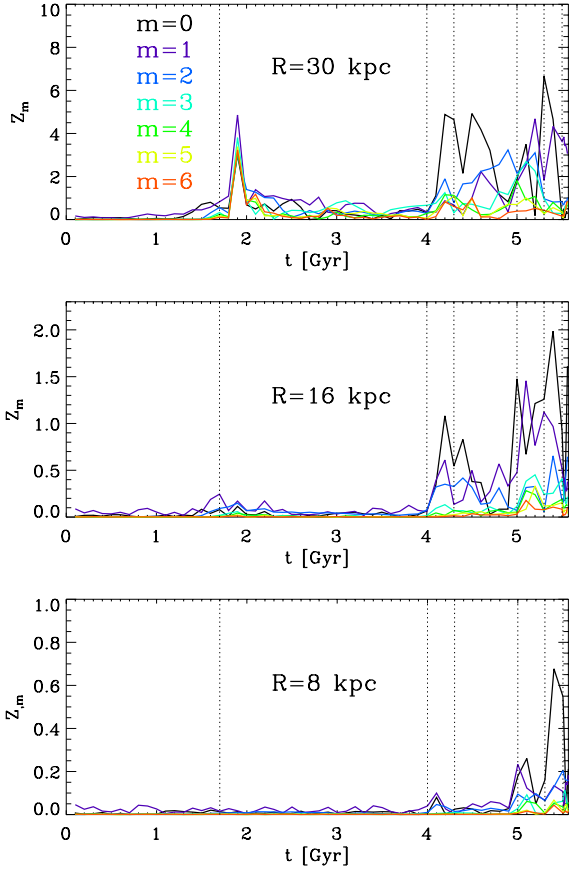}
\caption[]{Evolution of the amplitude of each Fourier mode measured for the mean vertical height of stars in model H2 at $R=22,16,8 \,\rm{kpc}$ respectively. The vertical dashed lines mark the times when Sgr crosses the $Z=0$ plane of the disc. The first pericentric passage of Sgr ($r_{peri}\sim 50 \,\rm{kpc}$) already strongly perturbs the outer-disc ($R\geq 16\,\rm{kpc}$). This is particularly important because features such as TriAnd, Monoceros are created as a result of cumulative interactions. The interaction with Sgr has appreciable effects on the vertical structure of the inner-disc only in the final phase of the satellite's disruption.}
\end{figure}

One way of quantifying the evolution of the vertical structure of the disc in time at a fixed radius $R$, is to use a Fourier decomposition of the disc:
\begin{equation}
\langle Z(R,\phi,t)\rangle= \sum_{m=0}^{\infty}\langle Z_{m}(R,t)\rangle e^{-im\phi}
\end{equation}
Using a set of three annuli positioned at $R=8, 16$ and $30\,\rm{kpc}$ respectively, we calculate the amplitude in each $m-$component of the Fourier series of the mean vertical height as a function of time. In Figure 17, we present the evolution of the amplitude in each mode as a function of time. We note that over the lifetime of the disc, the amplitude of the different modes in the disc grows coincidently and amplifies itself with every pericentric passage of the Sgr dSph. Furthermore, the mechanism by which the Galaxy is being perturbed and its stars are being kicked up as Sgr orbits the Galaxy operates in an {\it outside-in} fashion, starting in the outer-region of disc gradually making its way to the inner region\footnote{This is fundamentally different to perturbations seeded by noise which also have amplitudes that are lower by one to two orders of magnitude compared to those observed in the outer disc \citep{chequers17}.}. This is expected given the orbit of Sgr, which makes closer pericentric passages as dynamical friction brings the satellite to its present-day position in phase-space. Throughout the lifetime of the disc, the most dominant modes are the $m=0$, $m=1$ and $m=2$ distortions, however, as successive pericentric passages occur, these become amplified together with the higher order modes. Constrasting with the simulations of \cite{purcell11,gomez13,price-whelan15} we see that the previously ignored early pericentric passages are of high importance in shaping the disc as it is the repetitive aspects of the impacts with Sgr that dictate the final structure of the disc. At face value, this seems counter-intuitive as the first pericenter is far from the disc and the tides are probably not strong enough to perturb the disc.

The early interactions with Sgr also excite the emergence of transient spirals and strong radial motions. In Figure 18, we show the evolution of the surface overdensity of the disc. This is calculated by dividing the 2D surface density by the 1D radial density. This quantity is useful to visualise to what extent the Galaxy remains axisymmetric during its interaction with Sgr. We note that in the first couple of Gyrs of interaction with Sgr, the inner Galaxy (i.e. within the solar neighbourhood) does not deviate appreciably from axisymmetry whereas the outer parts show more prominant signs of departure from it through the emergence of transient spirals. In the last 2 Gyrs of evolution, however, strong spiral arms become evident and the Galaxy undergoes a buckling instability and forms a bar. Figure 19 shows the evolution of the mean radial motions in the disc. The order of the perturbation caused by Sgr on the disc can be as high as $100\,\rm{km/s}$ beyond $R>16\, \rm{kpc}$. Visual inspection between Figure 19 and Figure 16 reveal similar radial wavelengths for the perturbations showing that the motions are coupled \cite{d'onghia16}. The combined radial and vertical motions induced by the Sgr dSph on disc stars are non-negligible and certainly play an important role in shaping the structures such as the TriAnd clouds and Monoceros Ring. A more detailed look at the evolution of these features will be presented in Laporte et al. (in prep.), but for now, we will concern ourselves with identifying the driver which causes the vertical perturbations seen in our simulations.

\begin{figure}
\includegraphics[width=0.5\textwidth,trim=0mm 0mm 0mm 0mm,clip]{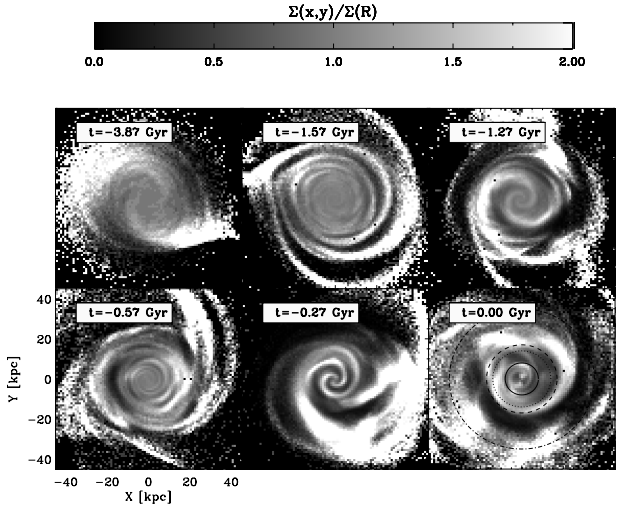}
\caption[]{Evolution of the surface overdensity for model H2. Each subsequent passage excites transient spiral arms, the last pericentric passage excites a bar instability which results in a bar of size $\sim4\,{\rm{kpc}}$ as revealed by a Fourier analysis as in Anathassoula et al. (2002)}
\end{figure}

\begin{figure}
\includegraphics[width=0.5\textwidth,trim=0mm 0mm 0mm 0mm,clip]{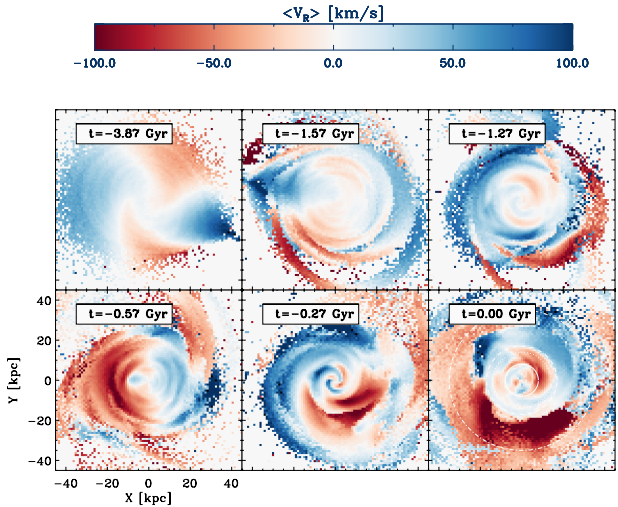}
\caption[]{Evolution of the radial stream velocities for model H2. These vary between $-100\,\rm{km/s}$ and $100 \,\rm{km/s}$. Sagittarius excites motions which are clearly coupled both in the radial and vertical direction. The radial motions correlate strongly with the locations of the spiral arms.}
\end{figure}

\subsection{Growth of the dark matter halo wake}

%Previous simulations of Sgr have generally neglected the infall phase of the Sgr dwarf from the virial radius. Contrary to conventional wisdom the very first pericentric passages of Sagittarius about the Milky Way have some real repercussion for the evolution of the outter disc, particularly in creating structures such as the TriAnd Clouds with the correct amplitude about the midplane. In fact, long before simulations of satellite disc encounters \cite{weinberg89} showed that as satellites' orbits decay as a result of dynamical friction, they also excite a wake in the halo which can penetrate all the way to the disc giving rise warps. We argued in both \cite{price-whelan15, gomez16} that these would be particularly important due to the combination of torques from the dark matter halo wake and tidal interactions.

\begin{figure}
\includegraphics[width=0.5\textwidth,trim=0mm 0mm 0mm 0mm,clip]{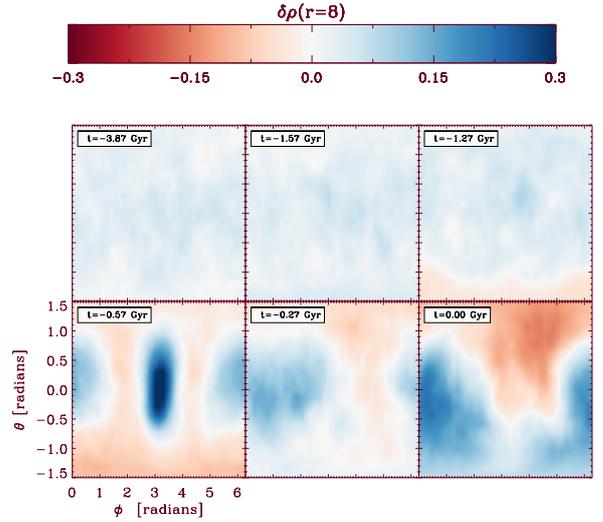}
\caption[]{Evolution of the wake of the dark matter halo at $r=8\,\rm{kpc}$. No dipole is seen until $t\sim 4.3 \rm{Gyr}$ which coincides with the time when the inner disc is being excited (see Figure 17) as Sgr makes a pericentric passage. Confirming the outside-in process by which the vertical oscillations of the disc are being seeded by the interaction with Sgr. }
\end{figure}

\begin{figure}
\includegraphics[width=0.5\textwidth,trim=0mm 0mm 0mm 0mm,clip]{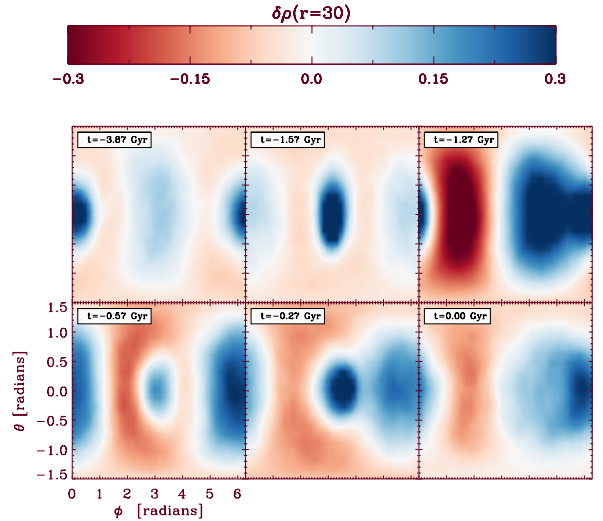}
\caption[]{Evolution of the wake of the dark matter halo at $r=30 \,\rm{kpc}$. A strong dipole changing in phase as Sgr orbits around the MW is visible coinciding with the times when the outer disc is being excited.}
\end{figure}

One explanation for the importance of early interactions despite Sgr's large perictenter is the mediation provided by the MW's dark matter halo. Using linear perturbation theory, \cite{weinberg98} was able to show that the formation of vertical perturbations could be excited by massive satellites as a natural consequence of dynamical friction which excites a dark matter wake inside a host halo which can penetrate in the inner region of the halo and affect the disc. \cite{vesperini00} also studied to what extent these dark matter halo wakes could be excited and found that even low velocity fly-bys may be able to excite wakes inside galaxies. In fact, using a cosmological simulation, \cite{gomez15b} was able to show that Monoceros-like features could even be excited by fly-by interactions of a subhalo with low velocities giving rise to a density wake in the host halo which translates itself into warping onto the disc. This was demonstrated by identifying the existence of a dipole feature in the dark matter overdensity distribution.  However, the simulation showed a dominant regular $m=1$ distortion in the mean vertical height of the disc, whereas the MW disc structure is more complex given the existence of Monoceros and other overdensities such as A13 and TriAnd which are both separated by $\sim 10 \,\rm{kpc}$. 

In our setup, the perturbing galaxy is Sgr and we want to find out whether it is also capable of exciting a dark matter wake. We proceed as in \cite{gomez15b} to calculate the overdensity of dark matter on a grid in spherical polar coordinates. The binsizes are chosen such that $(\Delta r,\Delta\phi, \Delta\theta)=(1, \pi/20, \pi/20)$, where the units are $\rm{kpc}$ for distance and radians for angles. The procedure calculates  a local density $\rho_{loc}$ by selecting all the DM particles that fall within a sphere $r_{sp}=3r\rm{sin}(\Delta\phi/2)$ at every grid point. The overdensity at every grid point is then calculated by subtracting the local density by the average density $\langle\rho\rangle_{|r}$ calculated within galactocentric shell of radius $r\pm r_{sp}$ as $\delta\rho(r,\theta,\phi)=\rho_{loc}/\langle\rho\rangle_{|r} -1$.

Figure 20 and 21 show the evolution of the dipole created in the dark matter density as the Sgr dSph orbits the MW at three different radii, $r=30, 8 \,\rm{kpc}$. We see that during the first passage across the disc, the DM halo already reacts at $r=30\,\rm{kpc}$ with overdensities reaching already 30 percent. Note that Sgr has a pericentric distance of $\sim 50 \,\rm{kpc}$ during its first passage, thus at those distances its tidal field is too weak to perturb the disc directly, leaving the wake as the only possible agent. As Sgr's orbit sinks closer into the halo, the wake is excited deeper at smaller radii and in the last 0.5 Gyr even down to $r=8 \,\rm{kpc}$. 

\subsection{Torque evolution: transitioning from host dominated regime to perturber domation}

We have identified a strong dipole inside the MW halo due to the dark matter wake that is naturally excited by Sgr through dynamical friction \citep{weinberg89}. However, Sgr in our model H2 run starts out quite massive with $M\!\sim 10^{11}\, \rm{M_{\odot}}$ and thus its tidal field in the vicinity of the disc is expected to be of some importance too. It is thus imperative to check the separate individual contributions from tidal interactions associated directly with Sgr's dark matter halo and those from the response of the MW's own dark matter halo. We proceed in a fashion similar to \citep{gomez15b} by taking a ring of 2000 test particles uniformly distributed at a radius of $R=16\,\rm{kpc}$ about the midplane of the disc and calculate the torques on the ring from the satellite\footnote{For Sgr, we selected all its originally associated dark matter particles within a spherical radius of $r=200 \,\rm{kpc}$, thus in calculating the torque we also take into account the DM stream that is left as a result of stripping too.} and the dark matter halo (for which we select particles between $0$ and $40 \,\rm{kpc}$). This is given by:

\begin{equation}
\tau^{\rm{shel}l}_{\rm{DM}}=\sum\limits^{2000}_{i=1}\mathbf{r}_{i}\times \mathbf{F}_{shell}
\end{equation}

Figure 22 shows the strength of the torque component perpendicular to the ring's angular momentum of each entity (MW halo wake\footnote{Our method is valid because of the symmetry of our setup. The force within a spherical shell of matter is zero by Newton's law. Because the MW halo is modelled as spherically symmetric, any force contribution within a spherical shell will come from the response of the MW halo to Sgr which we broadly call the wake. Our approximate calculation would not be valid if applied to cosmologically formed triaxial halo, as one would be need to make a clear distinction between torques from the intrinsic shape of the halo and those associated with the wake excited by an orbiting satellite.} and Sgr) normalised to that the maximum peak of the torque associated with the DM halo as a function of time. We note that the DM halo wake is excited at the first pericentric passage of Sgr. The vertical lines correspond to the times of pericentric passage of Sgr. %(note these do not always coincide with the times Sgr crosses the midplane). 
As the satellite reaches apocentre, another weaker peak emerges. During second approach another peak arises but with some delay, however during the third pericentric passage comes a transition where both the halo and Sgr contribute equally to the torque on the disc. After third pericentric passage, the roles are reversed and Sgr alone takes over the forcing of the disc as can be seen from its dominating signal during fourth approach. We thus conclude that the emergence of the vertical perturbation of the disc in the MW are intimately related to both its own dark halo and the Sgr dSph and are {\it distinct} in time and this proceeds as a {\it two-phase process}: one DM halo wake dominated regime followed by a later tidally dominated one.

\begin{figure}
\includegraphics[width=0.5\textwidth,trim=0mm 0mm 0mm 0mm,clip]{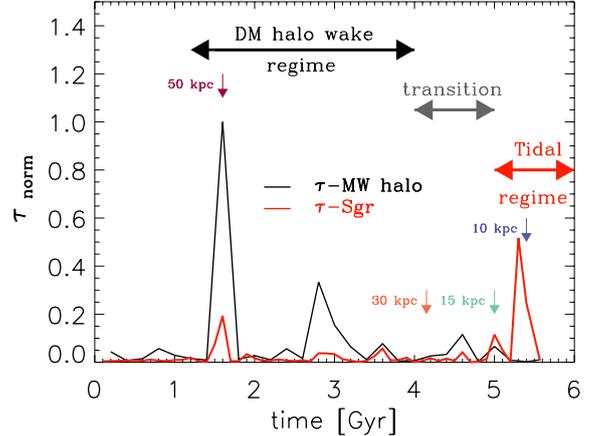}
\caption[]{Evolution of the torque from the MW dark matter halo and Sgr dSph on a ring of particles at $16\, \rm{kpc}$. The torques are normalised to the maximum of that of the DM halo. A two-phase mechanism is clearly evident. The vertical perturbations are mostly contributed from the torques from the DM halo wake on the disc and coincide with the first pericentric passage of Sgr. As Sgr reaches apocenter, the halo wake changes in phase and continues to excite disc material (see Figure 17 top panel). At further pericentric passages the torques from the halo wake continue to dominate. At the third pericentric passage, both the torques from the halo and tides from Sgr contribute equally to perturbing the disc, with the final phase of the torquing being completely dominated by the tides from Sgr. This confirms why the MW disc is particularly sensitive to the internal structure of Sgr as tides become increasingly important to sustain/amplify the vertical perturbations. }
\end{figure}

%\section{Three Musketeers: Sagittarius, the Magellanic Clouds and the Milky Way}
\section{The Response of the MW Galactic disc to both the LMC and Sgr}

In \cite{laporte16} we argued that the response of the MW to Sgr and the LMC are of comparable amplitude but may couple in non-trivial ways, thus implying the need to study their joint effect on the Galactic disc. The last Gyr of the evolution of the MW may be particularly affected by the LMC and Sgr simultaneously. This is because the LMC makes its first pericentric passage at the same warping the disc \citep{weinberg06, laporte16} as Sgr makes its final crossings of the Galactic disc. Thus it is important to study the signal imparted by both satellites as this could affect our earlier conclusions. We now run two simulations where we insert an LMC with progenitor mass $M_{200}\sim3\times10^{11}\,\rm{M_{\odot}}$ on an orbit presented in \cite{laporte16} 2 Gyr prior to the end of the original H1 and H2 simulations, in order to study the coupling of both satellites on the disc of the MW.

Figure 23 and 24 show maps of mean height and mean vertical streaming motions resulting from the coupling with the LMC. As noted in \cite{laporte16}, the LMC is indeed capable of modulating the structure of the Galactic disc in a non trivial way. This can be appreciated in maps of vertical mean height, particularly at large radii $R\geq 15 \rm{kpc}$, which coincides with the region where the Monoceros Ring is observed. We note that many of the features excited by Sgr alone remain, however, the influence of the LMC modulates their amplitude, both constructively and destructively interfering as function of azimuth. This is particularly noticeable in the the region $(R,l)=(15,135^{\circ})$. The inner region of the disc remains unaffected. %Moreover, the LMC excites further the amplitude of disc stars to extreme heights above and below the disc $|Z|\ge10\,\rm{kpc}$ in the very outer regions of the Milky Way 

{% A more complete mapping of the stellar disc and particularly its outer parts should have a strong potential to inform us on the MW's past interaction history.} 

%Certainly, if the MW was to be viewed as an external galaxy, some of its disc component could be mistaken for an accreted component. There is already indication in M31 through CMD fitting that outer features generally thought as part of the stellar halo may in fact belong to the disc \citep{ferguson05}. This is also translated in velocity space where the changes at large radii can amount to the order of  $\deltaV_{z}\sim10-25\,\rm{km/s}$ across the disc. While the LMC is a late protagonist in sculpting the disc, its gravitational impact on the disc is non-negligible showing interesting modulation.

\begin{figure}
\includegraphics[width=0.5\textwidth,trim=0mm 0mm 0mm 0mm,clip]{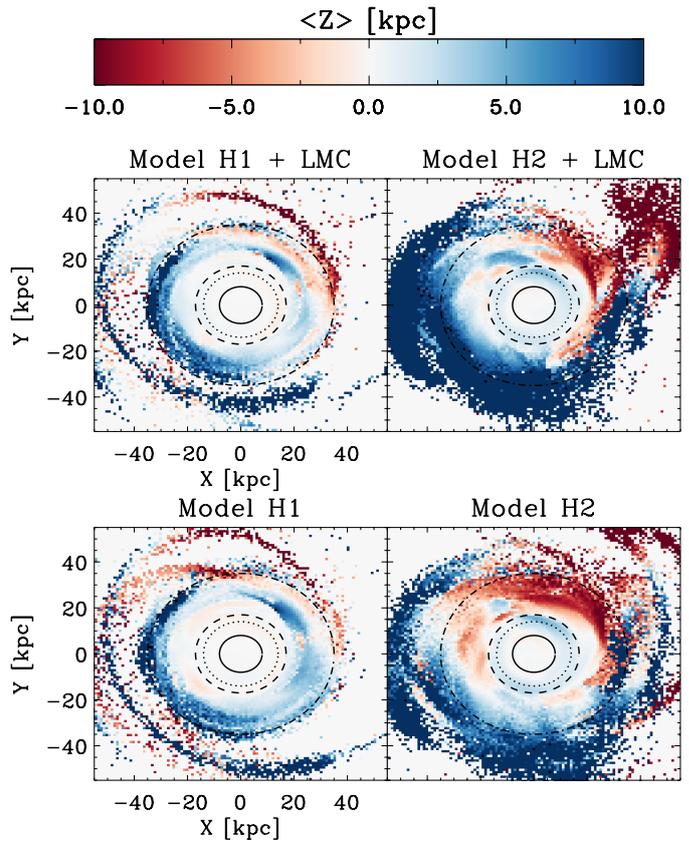}
\caption[]{Mean height maps of the disc for the LMC coupled runs and MW+Sgr runs. Many of the salient features of the individual Sgr-MW runs remain, however with a remarkable modulation of the mean height of the disc due to the coupling with the LMC through constructive and destructive interferences. These module the disc structure up to an order of $\Delta Z\sim3-5\,\rm{kpc}$ for $R\geq15\,\rm{kpc}$ }
\end{figure}

\begin{figure}
\includegraphics[width=0.5\textwidth,trim=0mm 0mm 0mm 0mm,clip]{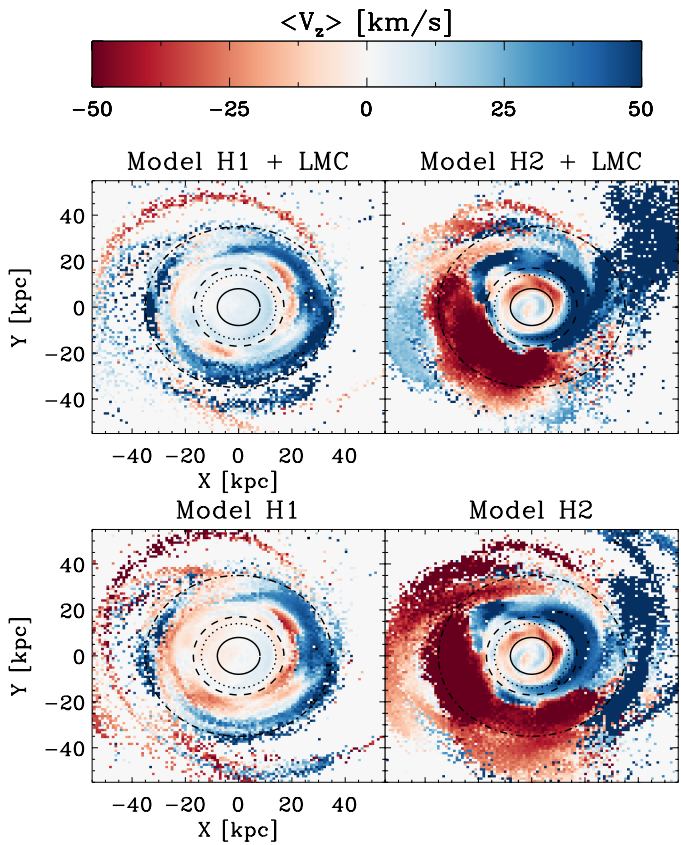}
\caption[]{Mean vertical streaming motion maps of the disc for the LMC coupled runs. Similarly to the mean height maps, we note that most of the structures excited by the Sgr-only runs remain qualitatively intact. The LMC only modulates those through its coupling through velocity changes of the order $\Delta V_{z}\sim10-25\,\rm{km/s}$ visible above $R\geq15\,\rm{kpc}$ when compared with the MW+Sgr runs. }
\end{figure}

\subsection{Implications for the HI warp of the Milky Way}

In \cite{laporte16} we suggested that the LMC alone might not be able to warp the HI disc of the Milky Way on its own and that possibly the interaction with Sgr may help lift material to higher heights than reported in our older simulations around $l\sim270^{\circ}$. Our simulations here do not model gas dynamics here and a direct comparison with the data from \citep{levine06} should not be made. However, we can still take a look at the outcome of the structure of the mean height of stars about the midplane around an annulus at a Galactocentric distance of $R\sim22 \,\rm{kpc}$ to illustrate the coupling of the LMC and Sgr on the structure of the stellar disc and contrast it with isolated runs (MW+Sgr and MW+LMC). This is shown in Figure 25 where we compare the mean vertical height of the stellar disc from \cite{laporte16}, our model heavy concentrated Sgr model H2 run, and their combination at a Galactocentric radius of $R=22 \,\rm{kpc}$ as a function of galactic longitude. We notice that the combined response is not a linear combination of the separate contributions from each satellite. We notice that in some regions the changes in amplitude between the LMC model and combined Sgr+LMC models can be as large as $\Delta Z\sim4-5\,\rm{kpc}$. As expected, the shape of the gaseous warp is not reproduced in this current work. This is not surprising as our numerical setup (collisionless N-body simulations) is not designed to address this kind of problem which requires a treatment of collisional dynamics. However, it is interesting to observe that the coupling between the LMC and Sgr was able to lift stellar material to much higher heights, notably towards the region of $l\sim280^{\circ}$ where we noticed a discrepancy of $\Delta Z\sim2-3 \,\rm{kpc}$ in our LMC alone models of the HI warp, suggesting that an interplay between the LMC and Sgr may be important in setting the structure of the HI cold gas warp. Nonetheless, it should be noted that the structure of the HI disc has been derived using kinematically inferred distances \citep{levine06}. Such methods are uncertain and known to be affected by systematic uncertainties of up to a factor of 2 \citep{reid14}. Thus, this casts doubts as to whether the discrepancies between our earlier models on the effect of the LMC alone on the warp of the Milky Way \citep{laporte16} and the maps derived in \cite{levine06} were significant. For now, a careful study of this question will need to wait for more appropriate gas modeling and/or likely a re-evaluation of the structure of the HI disc.

\begin{figure}
\includegraphics[width=0.5\textwidth,trim=0mm 0mm 0mm 0mm,clip]{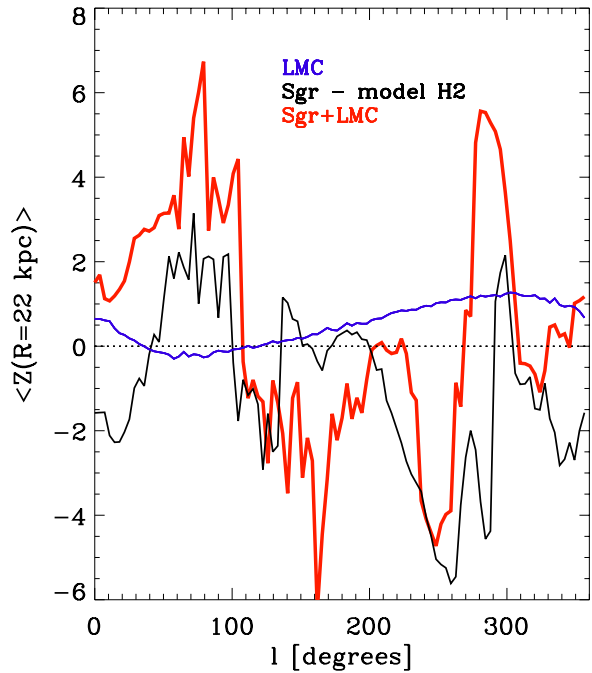}
\caption[]{Mean vertical height of the stellar disc at a Galactocentric radius of $R=22\,\rm{kpc}$ as a function of Galactic longitude $l$ for an isolated LMC, isolated Sgr model H2 and combined Sgr+LMC models shown as the blue, black and red solid lines respectively. The combined response of the MW disc is complex and not a simple linear combination of both separate contributions.}
\end{figure}

\section{Discussion}

\subsection{Limitations of our models}
Our experiment however idealised, shows much complexity, but it is important to note some caveats. We have considered in our initial conditions the present-day structure of the MW in equilibrium. In doing so, we are missing the time-evolution of the MW disc's potential which must have been different in the past. We expect that the potential of the disc must have varied as it grew in mass from $z\sim1$ to the present day and during the last 6 Gyr of its interaction with Sgr. During this time, the thin disc is still growing and this aspect is missed in the simulations. However, we can ascertain that the material seen in TriAnd and A13 could have initially come from pre-existing disc material some 6-8 Gyr ago. Detailed chemical abundance and age determination should clarify this. Furthermore, we have not explored different models for the MW's internal structure, but have concentrated ourselves on a fiducial MW model. Varying the structure of the disc (self-gravity and kinematic properties) would be another interesting exercise to study the dependence with the MW's response to Sgr. We also note that our experiment lacks a cosmological context under which the Galaxy as a whole is growing and interacting with its subhalos \citep{aumer13,grand17,ma17}. Nevertheless, it should be noted that cosmological simulations of MW-like halos are not yet able to capture the detailed accretion history of the MW (i.e. Sgr, MCs).

\subsection{Improvements in our model}
We have presented a series of models for Sagittarius focussing on the structure imparted on the Galactic disc. A particularly major improvement is that we have accounted for the infall of the Sgr dSph from the virial radius, whereas previous attempts have assumed initial conditions that were not realistic: a stable unperturbed disc with a truncated Sgr dSph at its Jacobi radius launched in the midplane \citep{purcell11,gomez13,price-whelan15}. In doing so these missed the previous two pericentric passages expected for Sgr, which we have self-consistently taken account of. The only other study considering similar types of initial conditions are the experiments of \citep{dierickx17}. Indeed, we find that accounting for the first pericentric passages of Sgr in the modeling is particularly important to account for the observed asymmetries at very large radii such as the Monoceros Ring and TriAnd overdensities. This is because although Sgr does not directly hit the disc during its first few pericentric passages, it still imparts a dark matter halo wake (which we have characterised as dipolar overdensities in the midplane in the halo of the MW) which affects the evolution of the stellar disc. Combined with the tides of Sgr, subsequent passages eventually force and amplify the vertical fluctuations further. Our Fourier decomposition show that this is exactly the case, with disc material already being excited to mean heights above $Z\sim2-3 \,\rm{kpc}$ at $R\sim30\,\rm{kpc}$ just after the first midplane crossing of Sgr's orbit. 

\subsection{Is Sagittarius the main culprit?}
Our experiments lend support for a common origin for Monoceros/GASS overdensities and many other low-latitude substructures such as A13 and TriAnd tied to the orbital history of Sagittarius. Other explanations for the existence of the Monoceros Ring in particular have been proposed. For example, the experiment of \cite{kazantzidis08} considers 6 encounters of low mass satellites with the thin disc and is able to qualitatively reproduce a Monoceros Ring out to the observed latitudes $|b|\sim30^{\circ}$. Such hypothetical interactions must have left a visible trace in the halo (either through streams or a phase-mixed component). However, the stellar halo of the MW is mostly dominated by the Sgr with a stellar mass of $3.7\pm1.2\times10^{8} \,\rm{M_{\odot}}$ between $1$ and $40\,\rm{kpc}$ \citep{bell08}. This constitutes about half of the mass from the reconstructed Sgr stellar mass of \cite{ostholt10} which is now locked in the stellar halo. Furthermore, cold streams from disrupted lower mass satellites are not observed in the vicinity of Monoceros \citep{slater14}. Moreover, looking at the chemical composition of the MW stellar halo, the $[\alpha/\rm{Fe}]-$poor knee falls within the range expected for massive known satellites such as the LMC and Sagittarius \citep{deboer14}, supporting the idea that the MW stellar halo predominantly formed from the accretion of massive objects. Additionally, kinematics from the stellar halo also suggest that the MW has had a particular quiescent accretion history \citep{deason17}. Thus having Sgr dSph as the main culprit behind the formation of the observed vertical perturbations seems to be supported by the observational constraints from the spatial, kinematic and chemical structure of the MW stellar halo.

From the cosmological simulations side, using the Auriga simulations \cite{gomez16} shows that multiple interactions with low mass satellites do not perturb the disc to the extreme heights of TriAnd. Instead, discs that have interacted with a $M\geq10^{10.5}\,\rm{M_{\odot}}$ satellite do and the mean velocities predicted in the outer regions show higher streaming velocity variations close to ones in our experiment. One exception to this are low mass fly-bys such as the one presented in \citep{gomez15b}. However, such an interaction only produces an $m=1$ perturbation and no multiple corrugation in the disc. For the MW, it is already not clear how many satellites have merged since $z\sim1$ with the Milky Way, if any \citep{ruchti15}, but there may still be a possibility for an ex-situ disc in the Milky Way \citep{gomez17}. Thus, despite its limitations, our experiment greatly improves on past attempts at modeling the interaction of Sgr with the disc and it is remarkable that it is able to make predictions in agreement with many observables of the MW disc as well as capturing the relevant physics behind the evolution of the outer-disc. At present, no cosmological simulation has been able to capture an accretion history that reflects that which is already being observationally witnessed for the Milky Way. This makes the scenario that Sgr was the main culprit behind the stellar disc structure all the more compelling. %Given that some of our Sgr models are able to capture all the observed distant features of the disc.

\subsection{What can outer disc structures tell us about Sgr?}

In our experiments, we note that the structure of the disc, particularly the Monoceros Ring is sensitive to mass loss history of the Sgr. This is reflected in the central density of the progenitor galaxy which we have varied. Thus it seems that the Monoceros Ring could potentially be used as a constraint on the internal mass profile of Sgr dwarf spheroidal's progenitor. Typically studies of the Sgr stream only consider varying the mass of the progenitor \citep{penarrubia10,law10,gibbons16} but not so much the internal mass profile. We caution that a proper exploration of studying the Sgr stream should take this larger parameter space into account as the disc is particularly sensitive to the orbital mass-loss history of Sgr. This is due to the fact that most of Sgr's mass gets stripped from infall to the present-day and the disc becomes particularly sensitive to remaining bound mass of the satellite. Using model H2, we showed that the Monoceros Ring gets excited in the last pericentric passage of Sgr which is the regime where the tidal torques from Sgr dominate (see Figure 22).  

Our work on the vertical structure of the Galactic disc lends additional support for a massive Sgr dSph progenitor. Previous studies have argued for a massive Sgr dSph progenitor starting as early as \cite{jiang00}. Fitting the stream alone has typically been difficult to argue for a massive Sgr progenitor \citep{johnston95,helmi01,helmi04,johnston05,law10, penarrubia10}. However, there exists other independent lines of evidence that have hinted towards a more massive progenitor. For example, \cite{ostholt10} have re-assembled the mass in the stream of the Sgr dSph and found that its total stellar mass would imply a dark halo progenitor of mass $\sim 10^{11}\,\rm{M_{\odot}}$ according to predictions from galaxy formation models \citep{Guo2010} and expectations from abundance matching \citep{Behroozi2012, Moster2013}. 

Moreover, \cite{gibbons16} have presented some N-body models of the stream for different progenitor masses, focusing in particular to the velocity dispersion in the streams. They favour a more massive progenitor with $M\geq 6\times10^{10} \,\rm{M_{\odot}}$. Their work hints towards a more massive progenitor. However, their disruption model is not fully realistic as it involves a discontinuity in the potential and time integration they consider. Furthermore, their experiments relies heavily on the ad-hoc dynamical friction formula of \cite{Chandrasekhar1943}. It will remain to be seen if such a massive progenitor with the right orbit in a self-consistent host potential can reproduce all the available constraints on the stream {\it and the disc}, as more observational data is being gathered \citep{sesar17}. Our models provide additional support for the need for a massive Sgr dSph from a different line of evidence.

\subsection{Did the Milky Way disc really form quiescently from $z\sim1$?}
 The existence of a thin disc, has led to generally think of the MW as an isolated disc galaxy which evolves through secular evolution to its present-day from $z\sim1$ to $z\sim0$ \citep{freeman02}. The accretion history of our Galaxy has been the focus of a number of studies trying to decypher the last important merger event of the Galaxy \citep{ruchti15}, finding that our MW has had a rather quiescent merger history.  However, our study shows that this picture may be misleading in the interpretation of the upcoming wealth of data from the {\it Gaia} as we show that the Sgr dSph is capable of strongly perturbing the MW disc and could be a potential agent in setting its present-day structure, affecting its star formation history, and chemical evolution. The field of streams is already a testament that accretion is still happening onto the Milky Way \citep{belokurov06}. More importantly, the Sgr dSph makes up the majority of the stellar halo within $\sim40\,\rm{kpc}$ \citep{bell08,deason11a}. Although the stellar mass locked into the stellar halo is insignificant to the mass budget within the disc, it is the invisible matter associated with Sgr's orbital history that is relevant to the evolution of the disc. We show that the MW's dark matter halo is reactive to Sgr's presence through the wake evolution and that the tides associated with Sgr in the last phases of disruption are non-negligible. Looking at the merging timescales from our simulations, Sagittarius must have been accreted between 8-6 Gyr ago, which would correspond to accretion times between $z\sim1$ and $z\sim0.7$ which is the time window during which the thin disc of the MW formed. If this is true, then Sgr must have been an important agent in shaping the chemical structure of the disc \citep{hayden15} and gradually flaring the different thin disc stars \citep{minchev15}. Given that we are lucky to be living in the MW and map both its smooth and lumpy structure, we may be able to uncover part of its formation history so as to further constrain its cosmological accretion history {\bf}. For now, our models show much promise and it will be imperative to tie both communities working on the galactic disc and Sgr streams in order to uncover the recent accretion history of the MW, for which Sgr dSph is the most important external contributor to setting the present-day structure of the Galaxy.

\subsection{Signatures from the LMC and Sgr dSph}

Finally, we studied the coupled response of Sagittarius with the LMC. In \cite{laporte16} we argued that the two satellites separately impart signals that are comparable in terms of the mean height of the disc. We confirmed our hypothesis, however, the main culprit behind the formation the vertical oscillations of the disc is still the Sgr dSph. We note that the coupling is only visible because of the large mass associated with LMC progenitor ($M_{halo}\sim 3\times 10^{11} M_{\odot}$). In \citep{laporte16} we showed that the LMC on a first infall produces a strong dipole in the MW halo with overdensity strengths as high as 50 percent at a distance of $\sim 30 \,\rm{kpc}$. Studying the kinematics of tracers in the halo in different quadrants of the sky (Garavito-Camargo et al. in prep.) may confirm this is case and further motivate studies on the combined effect of the LMC and Sgr on the disc. 

The models presented here put a lot of the multiple low surface brightness features of the disc under a common umbrella. These include the north-south asymmetries in physical space \citep{widrow12} but also the associated vertical motions of the disc which were also observed by RAVE and LAMOST \citep{williams13, carlin13}, GASS/Monoceros Ring \citep{newberg03,slater14,morganson16}, but also the existence of Cepheids at the opposite side of the Galaxy \citep{feast14} which have been interpreted to be associated with the flare (note that these could also belong to disc oscillations as the two are related - see figure 15). In fact, at the distances measured for these Cepheids ($R\sim15\,\rm{kpc}$), the disc strongly flares with $\sigma_{z}\sim2-5\, \rm{kpc}$.

%We have not explored different models for the MW's internal structure, but have concentrated ourselves on a fiducial MW model. Varying the structure of the disc would be another interesting exercise to study the dependence with the MW's response to Sgr. We leave this to a future contribution. We note that both Sgr and the LMC inflict strong perturbations on the internal structure of the MW's halo. Not only does this have important repercussions on the disc as we have seen here, but it should also affect the kinematics of the stellar halo as well as tracers such as globular clusters beyond $r=20 \,\rm{kpc}$. Which should be visible in measurements of the anisotropy parameter in different quadrants of the sky.  offers the opportunity to study a number of questions related to the MW disc dynamics, stellar halo and dark matter in the MW. 

Our work has several implications for the Milky Way and offers the opportunity to study a number of questions pertaining to its formation, evolution and dark matter. For example, to what degree can these perturbation affect our inferences on the local dark matter density in the solar neighbourhood and or across the disc? Is it possible to treat the transient response of the MW disc to Sgr using linear perturbation theory \citep{weinberg98,monari15}? What are the different Milky Way models which allow a massive Sgr progenitor and can we constrain its internal mass distribution using streams and the structure of disc \citep{gibbons16, dierickx17}? The presence of Sgr and the LMC must have important impacts on measuring the shape of the MW host potential \cite{veraciro14,gomez15}, to what extent will this impact the kinematics of stellar populations in the smooth halo of the MW (Garavito-Camargo in prep.)? Could the Sgr dSph have excited the bar in our Galaxy? We find many examples both in our realisations and past ones \citep{purcell11}. 

\section{Summary \& Conclusion}

The vertical disturbances in the MW within the solar neighbourhood, the Monoceros Ring and other overdensities such as TriAnd can {\it all be understood as part of the interaction of the disc with the Sgr dwarf galaxy}, with the condition that the progenitor was more massive than previously assumed. In summary, we show that:

\begin{enumerate}
\item As the Sgr dSph interacts with the MW, {\it the disc ripples and flares}. In this sense, Monoceros, A13 and TriAnd are not rings, but arching overdensities which are associated with oscillations of the disc. These are not symmetric in azimuth.

\item These features are sensitive to the mass of Sgr and particularly its orbital mass-loss history which is intrinsically related to its internal mass profile, characterised by its central density. This dictates how stripping proceeds, the timing and distance of disc-crossing impacts and how much bound mass remains. We show that a Sgr mass model with $\sim6\times10^{10} \rm{M_{\odot}}$ on the mean of the mass-concentration relation struggles to reproduce Monoceros and cannot excite material out to the known distances of TriAnd.

\item We presented a high-concentration model for Sgr dSph with a progenitor masses of $10^{11} \,\rm{M_{\odot}}$, which is capable of reproducing semi-quantitatively the observed features in the solar neighbourhood from \citep{widrow12}, the spatial extent in latitude of the Monoceros Ring with mean vertical proper motions $-1.5 \leq\mu _{b}/\rm{mas/yr}\leq 1.5$ that are consistent with observations \citep{deboer17} as well as the spatial location of the TriAnd overdensities $(l,b,d)$ with $R\sim30$ and $Z\sim -8 \,\rm{kpc}$ and the spatial location of the Cepheids from \cite{feast14} at $R\sim15\,\rm{kpc}$ on the opposite side of the Galaxy.

\item We demonstrate that the origin and evolution of these vertical perturbations comes from a two-phase scenario: The torques from the MW dark matter halo wake (as a result of dynamical friction) are crucial to excite the vertical perturbations in the disc for most of the interaction between Sgr and the disc. This can bring existing material in the outer disc out to $Z\sim10\,\rm{kpc}$ from the midplane easily, already during the first pericentric passage of Sgr. We show that the tides from Sgr are negligible in the first three pericentric passages are negligible to the effect of the MW dark matter halo. The torques from the tides from Sgr become important only towards the ends of the interaction during the last two pericentric passages.
\item Our Fourier decomposition shows that the vertical perturbations from a satellite infall such as Sgr are excited outside-in, leaving the inner disc intact at first and gradually penetrating further into the disc as the satellite sinks to the center. 
\item Sgr is the most important external contributor to shaping the structure and kinematics of the MW disc within the last 6-8 Gyr. This coincides with $z\sim1$ which is when the thin disc formed. The LMC is another important late perturber of the MW stellar disc and its interaction couples with that of Sgr in the last 2 Gyrs.
\item The signatures of the two satellites are unique: Sgr corrugates the disc whereas the LMC produces a large m=1 warp. The combination of the two modulates the response of the stellar disc but leaves many of the features excited by Sgr qualitatively intact, thus it may be possible to gain insight about their interactions from mapping these features. 

\end{enumerate}

The broad agreement of our model with the many features observed at the stellar-halo / disc boundary of the MW make it the most competitive model to explain the common existence of all these features as originating from the disc. Further observational campaigns combining photometry and spectroscopic analysis of elemental abundances should help us dissect the structure of the outer disc. The specific late accretion history of the Milky Way in the last 6-8 Gyr (with Sgr and in the last 2 Gyr the LMC) will be of relevance to cosmological simulations which are used to interpret the formation history of the MW and also observational chemodynamical studies trying to understand the formation of the MW in the last 8 Gyr from $z\sim1$. It is clear that the Sgr dwarf spheroidal will have an impact on the mixing of stellar populations in the disc (Laporte et al in prep). Although cosmological simulations are currently missing representations of Sgr and the LMC, the future is bright as it is hoped that this will eventually be addressed within the near future with constrained N-body simulations and higher computational power. Finding signatures of the DM halo wake in the stellar halo of the MW (Garavito in prep.) will be challenging but a worthwhile effort to further confirm our two-phase scenario for the origin and evolution of vertical disturbances in the disc. Our models of the outer disc are realistic enough to be useful to inform and motivate searches and spectroscopic follow-ups for kicked-up disc stars with ongoing/upcoming complementary surveys such as CFIS \citep{ibata17,ibata17b}, the Dark Energy Survey (DES), APOGEE-II \citep{zasowski17}, WEAVE \citep{famaey17}, 4MOST  \citep{feltzing17}, DESI \citep{desi16}. Finally, we note that it has been suggested in the past that the Sgr stream could be used to constrain the dwarf's orbital history \citep{johnston99}. It may be that combined with the constraints on the structure of the disc, we might someday be able to uncover the history of the Galactic disc in the last $6-8 \,\rm{Gyr}$. The scenario presented here should be falsifiable through observational campaigns designed to map out larger unobserved portions of the outer disc in phase-space as well as elemental abundances and stellar ages.

\section*{Acknowledgments}
We thank Volker Springel for giving us access to the {\sc gadget-3} code. We thank valuable discussions with Benoit Famaey, Rodrigo Ibata, Keith Hawkins, Vasily Belokurov, Jorge Pe\~narrubia, Rapha\"el Errani. CL is supported by a Junior Fellow of the Simons Society of Fellows award from the Simons Foundation. This work used the Extreme Science and Engineering Discovery Environment (XSEDE), which is supported by National Science Foundation grant number OCI-1053575. We also acknowledge use of computing facilities at the Rechenzentrum Garching (RZG) and the Max Planck Institute for Astrophysics (MPA). KVJ's contributions were supported by NSF grants AST-1312196 and AST-1614743. N.G-C is supported by the McCarthy-Stoeger scholarship from the Vatican Observatory. CL dedicates this work to all the New York jazz cats and in particular his mentors Aaron Parks, Donald Vega for their continued support and Dr. Barry Harris for his teachings. 
\bibliographystyle{mnras}
\bibliography{master2.bib}{}

\appendix

\section{Sgr stream}
We note that our models perform at a reasonably good level in reproducing the extent of the stream on the sky, the ranges of heliocentric distances of the leading and trailing arms, as well as the line-of-sight velocities of the stream. It is important to note that having a fully self-consistent model for Sgr (i.e. with a live halo and no ad-hoc prescriptions for dynamical friction) is a daunting task. In fact, our models perform just as well as the careful live N-body experiments such as the one from \citep{dierickx17}. Although these authors are able to reproduce the full 6-D motion of the core from Sgr, their streams are not a better match than any of the models presented here. This reflects the complexity of the problem of self-consistently modeling the Sgr stream from infall of its progenitor to the present-day which so far has only been attempted by one study. With this in mind, we judge that while our models could be improved in the future to better match the streams and the structure of the disc, it is reasonable to already study the predictions from such simple yet reasonably accurate models on the structure of the Galactic disc.

%We note in passing that a detailed modeling of the Sgr stream in a fully self-consistent live potential is particularly difficult and given the uncertainties of the MW potential and shape we have not tried to find a perfect match with the final position and velocity of Sgr \citep{dierickx17}. However, it is expected that more detailed models would not significantly alter the conclusions from our work in terms of the salient features we are about to report here. 

\subsection{Sgr streams}
\begin{figure}
\includegraphics[width=0.5\textwidth,trim=0mm 0mm 0mm 0mm,clip]{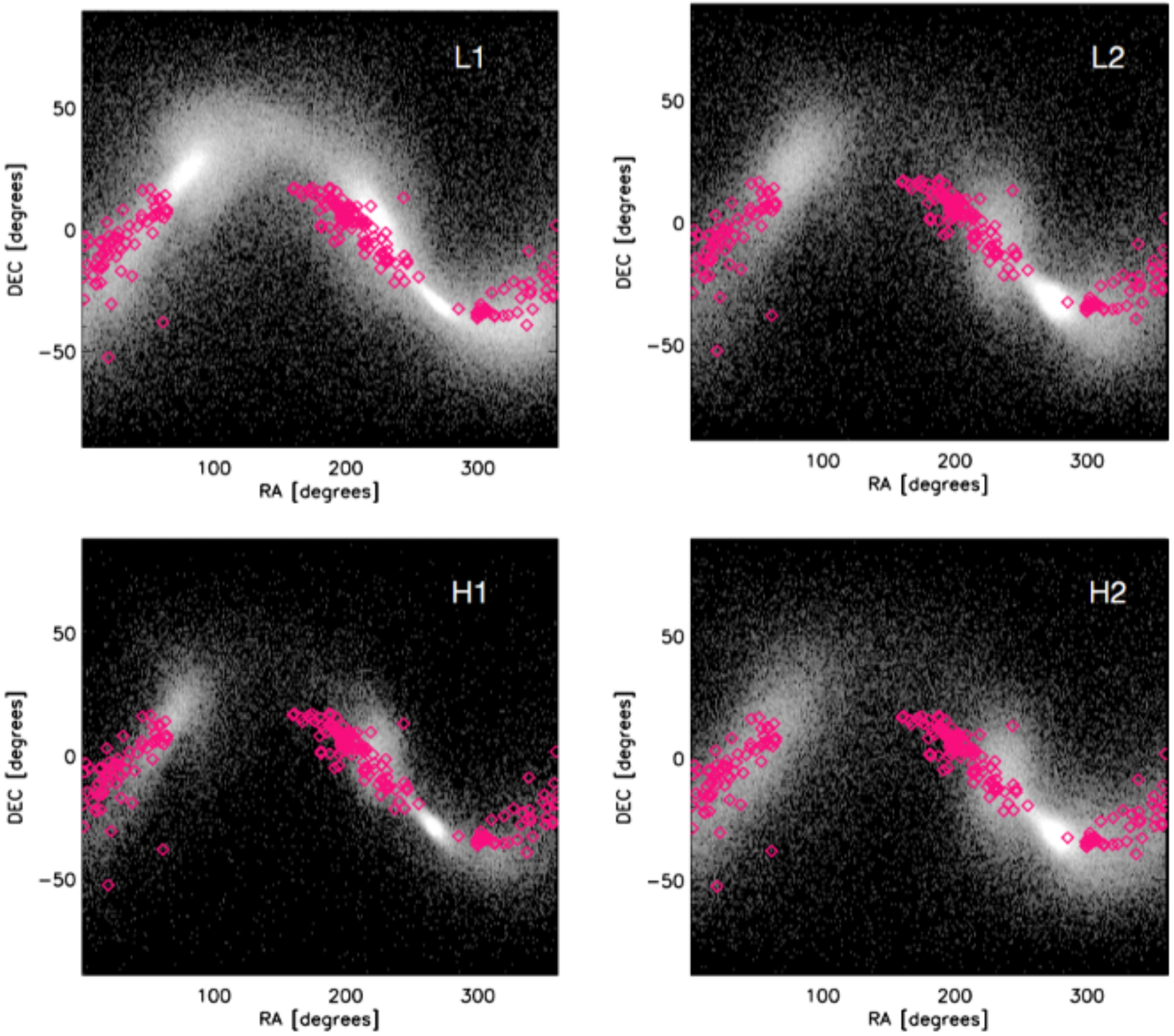}
\caption[]{Distribution of N-body stars particles for all four models (L1,L2,H1,H2) on the sky in $(ra,dec)$. The pink diamonds are the data from the 2MASS M-Giants from trailing and leading arms of Majewski et al. 2004. Overall the models are able to qualitatively reproduce the distribution of stars in the stream to a reasonable level. The simulations were not fine tuned to match the data.}
\end{figure}

\begin{figure}
\includegraphics[width=0.5\textwidth,trim=0mm 0mm 0mm 0mm,clip]{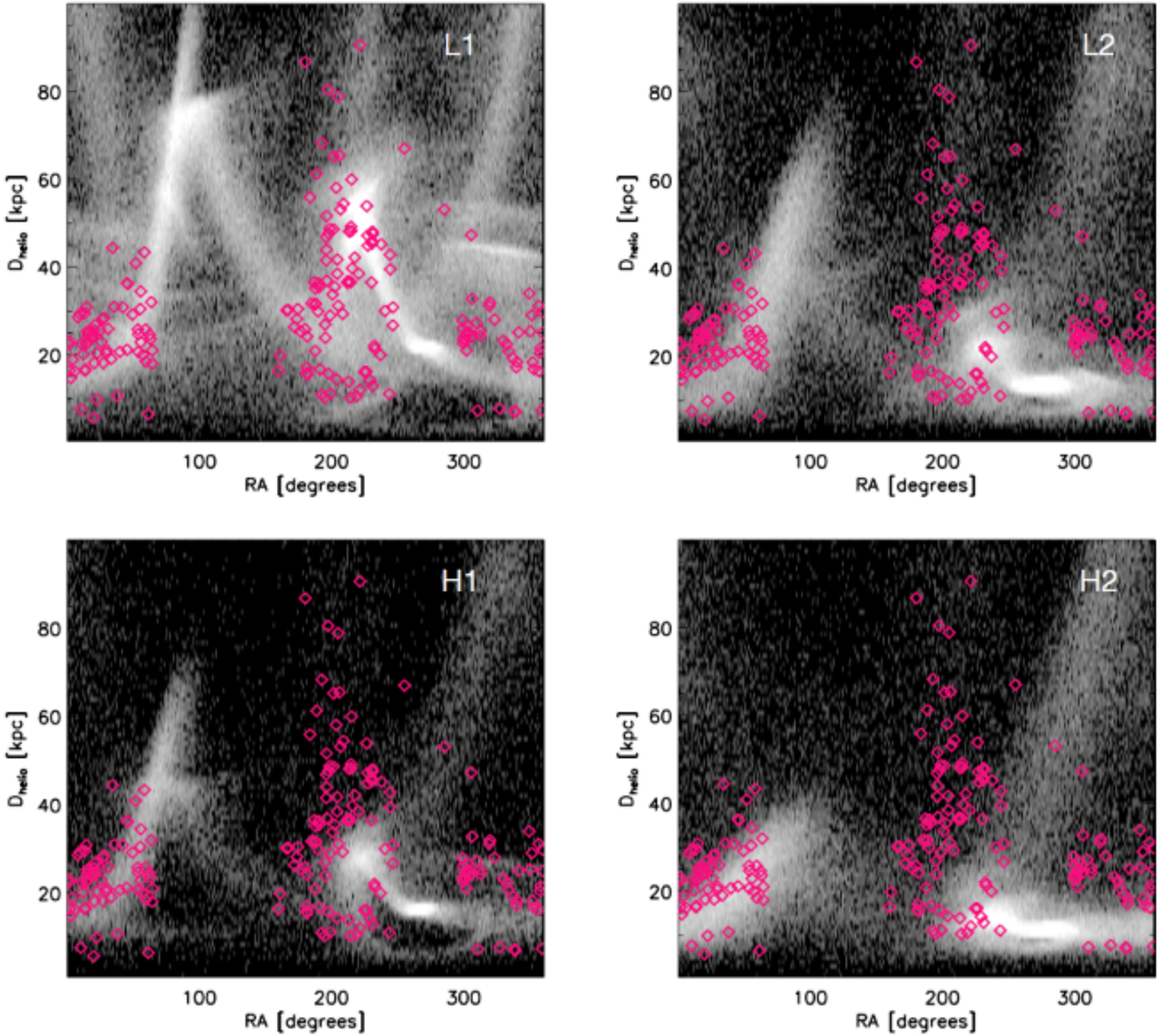}
\caption[]{Distribution of N-body stars particles' heliocentric distances for all four models (L1,L2,H1,H2)  as a function of right ascension. The pink diamonds are the data from the 2MASS M-Giants from trailing and leading arms of Majewski et al. 2004. We note that all models are able to produce streams at all measured heliocentric distances, though with different degrees in density. Given that the models were not fine tuned to match the data, we consider these as acceptable in terms of being analogues of Sgr interacting with the MW.}
\end{figure}

\begin{figure}
\includegraphics[width=0.5\textwidth,trim=0mm 0mm 0mm 0mm,clip]{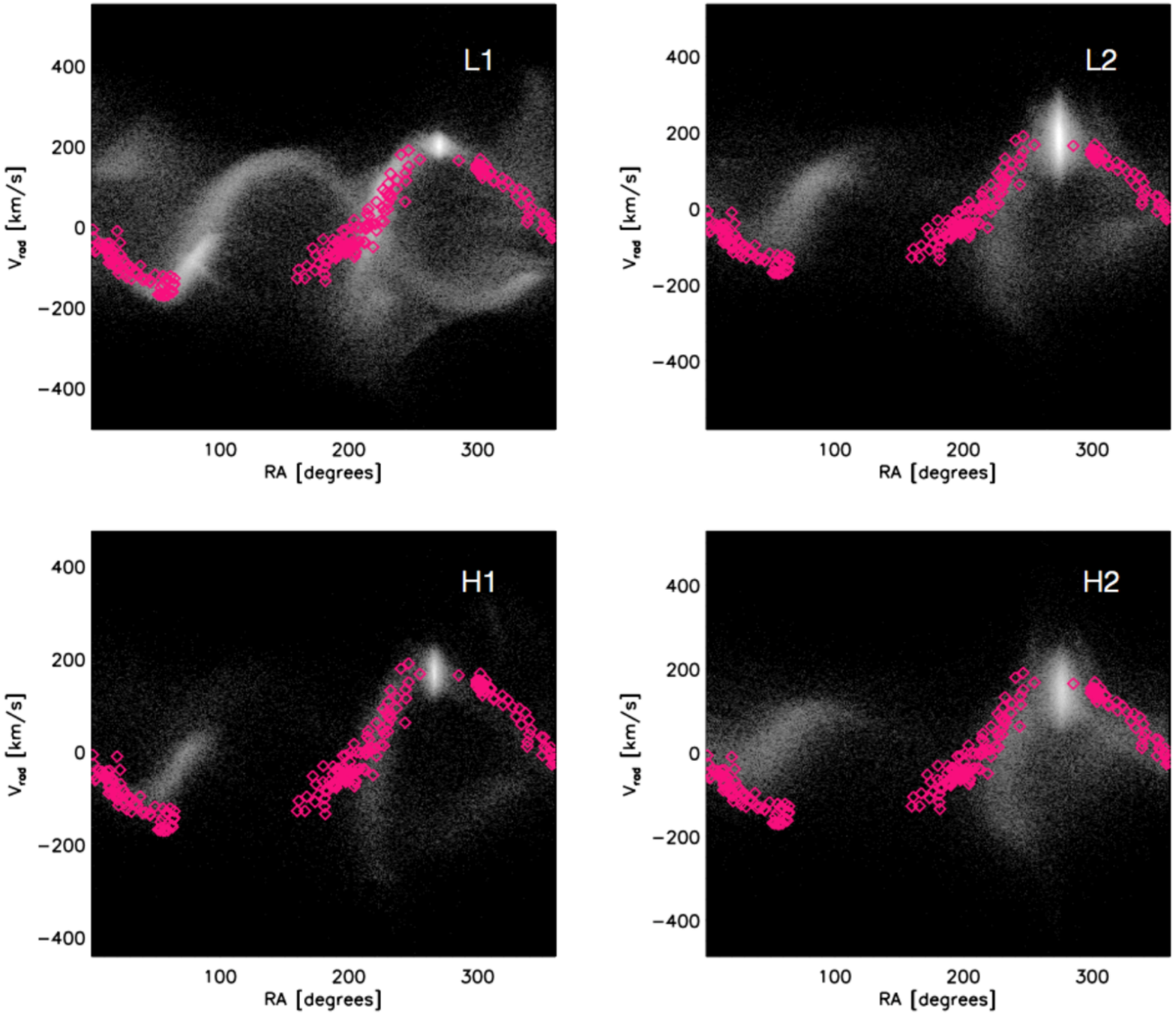}
\caption[]{Distribution of N-body stars particles radial velocities for all four models (L1,L2,H1,H2) as a function of right ascension. The pink diamonds are the data from the 2MASS M-Giants from trailing and leading arms of Majewski et al. 2004. Overall the models are able to qualitatively reproduce the distribution of radial velocities in the stream to an acceptable level of success. Note as in all previous models in spherical potentials  the difficulty of matching the tip of the leading arm.}
\end{figure}

Given the high masses we consider for the progenitor of Sgr, we checked whether the final central velocity dispersions of the remnant are in agreement with the accepted measured values \citep{frinchaboy12} which vary between 15 and 20 km/s. Using our N-body representation of the stellar parts, the L2 and H2 models have dispersions that are much too high, by a factor of two. However, we note that we have assumed Hernquist spheres for the stellar components of the live Sgr progenitor models. Given that Sgr is a dark matter dominated galaxy, the stars can be considered as tracers of the total potential. Thus it is possible for the given N-body representation to re-weight the original dark and star particles \citep{Laporte2012} such as to create new distributions functions in equilibrium with the underlying total original potential in the initial conditions. We do this using the machinery from \citep{Bullock2005} which has been generalised to fully triaxial systems by \citep{Laporte2013}. By calculating new distribution functions for the stellar component of Sgr following plummer profiles, we are able to bring the stellar velocity dispersion of the remnants down to 19 km/s which is within the measured values for Sgr. Thus we conclude that for all our progenitor total mass models, there exists a distribution function for the luminous component of Sgr for it to have a final velocity dispersion compatible with the measured value for the core of Sgr.

\label{lastpage}
\end{document}